\documentclass[pdflatex,sn-mathphys-num]{sn-jnl}
\usepackage{graphicx}%
\usepackage{multirow}%
\usepackage{amsmath,amssymb,amsfonts}%
\usepackage{amsthm}%
\usepackage{mathrsfs}%
\usepackage[title]{appendix}%
\usepackage{xcolor}%
\usepackage{textcomp}%
\usepackage{manyfoot}%
\usepackage{booktabs}%
\usepackage{algorithm}%
\usepackage{algorithmicx}%
\usepackage{algpseudocode}%
\usepackage{listings}%
\usepackage{upgreek}
\usepackage[mathscr]{eucal}
\usepackage{bbm}
\usepackage{array}
\usepackage{natbib}

\theoremstyle{thmstyleone}%
\theoremstyle{thmstyletwo}%

\theoremstyle{thmstylethree}%

\begin{document}

\title[Classical scattering matrix for hard and soft Bose-excitations in a non-Abelian plasma within the Hamiltonian formalism]{Classical scattering matrix for hard and soft Bose-excitations in a non-Abelian plasma\\ within the Hamiltonian formalism}


\author[1]{\fnm{Yu.A.} \sur{Markov}}\email{markov@icc.ru}

\author[1]{\fnm{M.A.} \sur{Markova}}\email{markova@icc.ru}

\author*[2,3]{\fnm{D.M.} \sur{Gitman}}\email{dmitrygitman@hotmail.com}

\author[1,4]{\fnm{N.Yu.} \sur{Markov}}\email{nyumarkov@gmail.com}

\affil[1]{\orgdiv{Matrosov Institute for System Dynamics and Control Theory},\\ \orgname{Russian Academy of Sciences}, \orgaddress{\street{Lermontov str., 134}, \city{Irkutsk}, \\ \postcode{664033}, \country{Russia}}}

\affil[2]{\orgdiv{P.N. Lebedev Physical Institute}, \orgname{Russian Academy of Sciences}, \orgaddress{\street{Leninskiy ave. 53}, \city{Moscow}, \postcode{119991}, \country{Russia}}}

\affil[3]{\orgdiv{Institute of Physics, University of S\~ao Paulo}, \orgaddress{\street{Rua do Mat\~ao, 1371},\\ \city{S\~ao Paulo}, \postcode{05508-090}, \country{Brazil}}}

\affil[4]{\orgdiv{Irkutsk State University}, \orgname{Russian Academy of Sciences},\\ \orgaddress{\street{Karl Marx str., 1}, \city{Irkutsk}, \postcode{664003}, \country{Russia}}}


\abstract{Within the framework of the Zakharov-Schulman approach, in close analogy with the methods of quantum field theory, the classical scattering matrix for the simplest process of interaction between hard and soft excitations in a quark-gluon plasma (QGP), is determined. The classical $\mathcal{S}$-matrix is defined in the form of the most general integro-power series expansion in the asymptotic values as $t\rightarrow-\infty$ of normal bosonic variables $c^{-\,a}_{\hspace{0.02cm}{\bf k}}(t)$ and $(c^{-\,a}_{\hspace{0.02cm}{\bf k}}(t))^{\ast}$, describing the soft gluon excitations of the system, and a color charge $\mathcal{Q}^{-\hspace{0.03cm}a}(t)$ of a hard particle. The first nontrivial contribution to this matrix is calculated. The quantum commutator of quantum field operators is replaced by the so-called Lie-Poisson bracket depending on the classical asymptotic variables. The developed approach is used to derive a general formula for energy loss of a fast color-charged particle during its scattering off soft bosonic excitations of QGP in the framework of the classical Hamiltonian formalism. For this purpose, the notion of an effective current of the scattering process under consideration is introduced and its relation to the classical $\mathcal{S}$-matrix is determined. With the help of the known form of the classical scattering matrix, the desired effective current is recovered, which in turn allowed us to determine the formula for  energy loss of the hard color particle. The rough estimates of energy loss at the 	order-of-magnitude level is provided and their comparison with the well-known results on the radiation and collision losses is performed.}





\maketitle

\section{Introduction}\label{sec1}

In \cite{Markov:2023rlz} we suggested a Hamiltonian theory for collective longitudinally polarized Bose-excitations (plasmons) interacting with a classical high-energy color-charged particle propagating through a hot quark-gluon plasma. For this purpose, we applied a general formalism for constructing the wave theory in nonlinear media with dispersion based on the classical Hamiltonian theory of systems with distributed parameters, proposed in due time by Zakharov \cite{Zakharov1971, Zakharov:1974}, developed later by Gitman and Tyutin \cite{Gitman:1990} for quantum fields and presented in detail on numerous examples of concrete physical systems in the reviews \cite{Zakharov:1985, Zakharov:1997} and in the monograph \cite{Zakharov:1992} (see also \cite{Krasitskii:1990, Markov:2020efa, Markov:2021nhe}).\\
\indent In the present work, as a concrete physical application of the Hamiltonian wave theory of quark-gluon plasma, we consider the problem of calculating the energy loss of ultra-relativistic color-charged particles passing through a hot QCD medium. As is well known, energy loss is one of the most important tools for diagnostics of the quark-gluon plasma in ultrarelativistic heavy-ion collisions \cite{ALICE:2022wpn}. In this paper, we are only interested in the leading hard thermal loop (HTL) contribution with respect to the strong coupling constant. This allows us to simplify the treatment and consider the high-energy massless particle as moving along a straight trajectory with a constant velocity
\[
{\bf x} = {\bf x}_{0} + {\bf v}\hspace{0.03cm}(t - t_{0}),
\] 
where ${\bf x}_{0}$ and ${\bf v}$ are the initial position and velocity of the hard particle, respectively. In the HTL-approximation, the Abelian part of the Compton scattering is suppressed, and only the dominant specific non-Abelian contribution survives. This is a standard approach in the field, which is reasonable in the high-energy regime.\\
\indent In spite of the fact that we assume the trajectory of a hard particle to be straight and its velocity to be constant, the particle under consideration loses energy due to the rotation of its color charge in an effective color space during the scattering off the soft gluon excitations of the quark-gluon plasma\footnote{\hspace{0.03cm}We specifically note that here the rotation of color charge determines only radiative energy loss. There is another important class of energy losses associated with direct collisions with thermal particles of the QCD medium. Such particle collisions also lead to the rotation of their color charges, as soon as a particle ``hits'' the field of the other particle at the collision point thereby resulting in the collision energy losses (see, for example, \cite{Kovchegov:1997}).}. The rotation of the color charge of the particle leads to the emission (absorption) of soft bosonic excitations. The most natural approach to obtaining an expression for energy loss is the method developed for the ordinary Abelian (electron-ion) plasma. A thorough discussion of this topic can be found in the monograph by Akhiezer {\it et al.} \cite{Akhiezer:1975}. It is only necessary to make a minimal generalization to the color degrees of freedom for soft and hard excitations in QGP.\\
\indent We should at once specify that we are interested in energy  losses of a hard particle in the region of momentum/energy transfer of the order of the so-called Debye mass, i.e. at $q\sim g\hspace{0.02cm}T\sim m_{D}$. In this region, the description that takes into account the screening effects and the presence of quasiparticle excitations of the quark-gluon plasma, is valid. Here, we are dealing with the hard thermal loop perturbation theory \cite{Blaizot:2002, Braaten:1990, Ghiglieri:2020}.
These losses are most likely of purely academic interest, since it is well known (see, for example, \cite{Jacobs:2006}) that for fast partons with energy $E\gg T$ (from tens to hundreds of GeV for RHIC and LHC conditions) the main part of the losses comes from the medium-induced gluon bremsstrahlung and from the direct elastic collisions with hard thermal particles. The latter are related to the scattering processes with momentum transfer to the medium particles of the order $q\gg g\hspace{0.02cm}T$, at which the quasi-particle description does not work anymore, since such phenomena are beyond the region of validity of the HTL perturbation theory.\\
\indent In fact, the main purpose of this work is to answer the question: is it possible, while remaining within the framework of the Hamiltonian wave theory alone, to obtain a formula for the energy loss of a fast color-charged particle during its scattering off soft collective bosonic (and, in general, fermionic) excitations of a non-Abelian plasma.The calculation of energy loss in this approach requires knowledge of the effective bosonic current for particles with integer spin or of the effective fermionic current for particles with half-integer spin, which are generated by the scattering of the particles off the collective waves of the medium or by the scattering of hard particles off each other. The latter determines the energy losses due to bremsstrahlung, while the former is due to the so-called spontaneous scattering processes. Thus, to obtain the required expression of energy loss, it is necessary to know the effective currents of bosonic or fermionic types associated with the scattering processes interesting to us.\\
\indent In our case to calculate this effective bosonic current, staying only within the framework of the Hamiltonian theory, we will use the expression for the so-called {\it classical scattering matrix}. The matrix was introduced for the first time by Zakharov \cite{10.1007/3-540-11192-1_38} for Hamiltonian wave systems and then was developed in the works of Zakharov and Schulman {\it et al.} \cite{ZakShu85, Zakharov:1988, Zakharov:2014}. However, in these works, the scattering matrix was determined, so to speak, only for the soft sector of excitations of physical systems. The sufficient universality of this approach allowed us to propose for the first time a method for constructing a classical $\mathcal{S}$-matrix for a highly excited strongly interacting system, such as the quark-gluon plasma coupling with hard color-charged partons. Due to the complexity of the problem, in this work we restricted ourselves to the simplest interaction process -- elastic scattering of one energetic particle off a plasmon. As is known, in the framework of quantum field theory (see, for example, the monographs by N.N. Bogolubov {\it et al.} \cite{Bogolyubov:1975ps, Bogolubov:1990}) the operators of bosonic and fermionic currents represent the so-called first-order radiation operators, which in turn are expressed through the variational derivatives of the quantum $S$-matrix. We suppose to apply these relations to obtain the classical bosonic and also  fermionic currents, where the classical $\mathcal{S}$-matrix in the spirit of Zakharov-Schulman approach will be used instead of the quantum $S$-matrix. \\
\indent The method of defining the effective bosonic current on the basis of the $S$-matrix has already been used in a number of works as an application to the problems of a hot QCD medium. For example, Jackiw and  Nair \cite{PhysRevD.48.4991} have used the bosonic current to derive high-temperature response functions for a non-Abelian plasma and the corresponding non-Abelian generalization of the Kubo formula. In the paper by Elmfors, Hansson, and Zahed \cite{PhysRevD.59.045018}, the formula relating the current and the $S$-matrix was used to simply derive the effective action for hard temperature loops.\\
\indent The paper is organized as follows. In section \ref{section_2}, we present an explicit form of the previously obtained in \cite{Markov:2023rlz}  the fourth-order effective Hamiltonian ${\mathcal H}^{(4)}$ describing the elastic scattering of a hard color particle off the collective longitudinal QGP excitations. The Lie-Poisson bracket, which is used within the canonical formalism and the corresponding Hamilton equations for the basic dynamical variables, are written out. A diagrammatic interpretation of the various terms included in the effective scattering amplitude, is given. In section \ref{section_3}, a general approach to the determination of the classical scattering matrix for the process of interaction of a hard color-charged particle with soft bosonic excitations of QGP, is presented. For this purpose, the so-called adiabatic hypothesis of switching off the interaction is used. Section \ref{section_4} is devoted to the explicit derivation of the $\mathcal{S}$-scattering matrix for the simplest case of the interaction Hamiltonian (\ref{eq:2q}) quadratic in the normal field variables and linear in a color charge. The scattering matrix is given here in the form of some integral operators relating the asymptotic in-variables  $c^{-\,a}_{\hspace{0.02cm}{\bf k}}$ and $\mathcal{Q}^{-\hspace{0.03cm}a}$ as $t\rightarrow -\infty$ with the asymptotic out-variables $c^{+\,a}_{\hspace{0.02cm}{\bf k}}$ and $\mathcal{Q}^{+\hspace{0.03cm}a}$ as $t\rightarrow +\infty$.\\
\indent In section \ref{section_5}, by analogy with quantum field theory, the classical $\mathcal{S}$-scattering matrix is constructed in the form of an exponential function rather than as the integral operator. For this purpose, the Lie-Poisson bracket in the new asymptotic in-variables $c^{-\,a}_{\hspace{0.02cm}{\bf k}}$, $(c^{-\,a}_{\hspace{0.02cm}{\bf k}})^{\ast}$ and $\mathcal{Q}^{-\hspace{0.03cm}a}$ is used. In section \ref{section_6}, we give a general expression for energy loss of a fast charged particle moving in a usual electron-ion plasma with a minimal generalization to the color degree of freedom. Section \ref{section_7} is devoted to the construction of an effective current of the hard color particle coupling with a high-temperature quark-gluon plasma. This current is defined in full analogy with quantum field theory -- in the form of the first-order radiation operator (current operator). The effective current thus found in the coordinate representation is then rewritten in the Fourier representation. In section \ref{section_8}, on the basis of the effective current obtained, the derivation of the final expression for energy loss, is presented. The contributions to the energy loss from the asymptotic scalar colorless $N^{-\hspace{0.03cm}l}_{\bf k}$ and color $W^{-\hspace{0.03cm}l}_{\bf k}$ components of the plasmon number density ${\mathcal N}^{\hspace{0.02cm}-\hspace{0.02cm}a\hspace{0.03cm}a^{\prime}}_{\bf k}$ are analyzed separately. Section \ref{section_9} discusses the calculation of an order-of-magnitude estimate of energy loss of a hard particle under two extremely opposite conditions: when a hot QCD plasma, in the asymptotic past, close to thermal equilibrium, and when it is in a strongly excited state.\\
\indent In the concluding section \ref{section_10}, we briefly summarize our findings  and outline possible ways of their generalization to the fermion sector of hard and soft excitations of the quark-gluon plasma. An explicit form of the effective gluon propagator and of the effective three-plasmon vertex in the hard thermal loop approximation, which we use in the paper, is given in Appendix.

\section{\bf Interaction Hamiltonian of plasmons and a hard color particle}
\label{section_2}

\begin{figure}[hbtp]
	\vspace{-0.2cm}
	\begin{center}
		\includegraphics[width=1\textwidth]{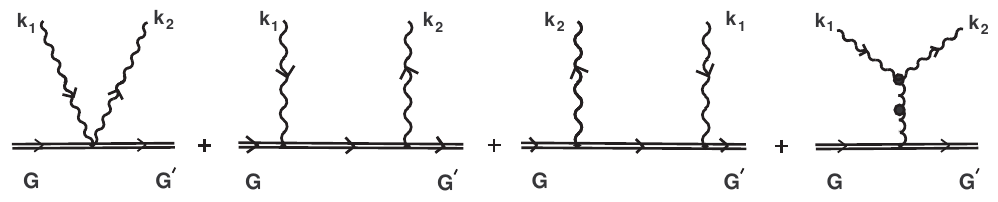}
	\end{center}
	\vspace{-0.2cm}
	\caption{\small Effective amplitude $\widetilde{T}^{\,(2)\,a_{1}\hspace{0.03cm}a_{2}\, a}_{\,{\bf k}_{1},\, 
			{\bf k}_{2}}$ for the elastic plasmon scattering process off a hard colored particle. The double line denotes the hard particle carrying a color charge $\mathcal{Q}^{\hspace{0.03cm}a}$ and the black dots denote the HTL summation}
	\label{fig1}
	\vspace{-0.3cm}
\end{figure}

$\quad\,$ For the convenience of further references, this section provides the necessary minimum information from \cite{Markov:2023rlz}. In particular, in this paper, an explicit form of the effective fourth-order Hamiltonian ${\mathcal H}^{(4)}$, which describes the elastic scattering of the collective longitudinal excitations (plasmons) off a hard color-charged particle, was obtained:
\begin{equation}
	{\mathcal H}^{(4)}_{g\hspace{0.02cm}G\hspace{0.02cm}\rightarrow
		\hspace{0.02cm} g\hspace{0.02cm}G} 
	=
	i\!\int\!d\hspace{0.03cm}{\bf k}_{1}\hspace{0.03cm}d\hspace{0.03cm}{\bf k}_{2}\,
	\mathscr{T}^{\hspace{0.03cm}(2)\hspace{0.03cm} a\,a_{1}\hspace{0.03cm}a_{2}}_{\; {\bf k}_{1},\, {\bf k}_{2}}\,
	c^{\ast\ \!\!a_{1}}_{\hspace{0.03cm}{\bf k}_{1}} c^{\hspace{0.03cm}a_{2}}_{\hspace{0.03cm}{\bf k}_{2}}
	\mathcal{Q}^{\hspace{0.03cm}a},
	\label{eq:2q}
\end{equation}
where the amplitudes $c^{\ast\ \!\!a}_{\hspace{0.02cm}{\bf k}}$ and $c^{\hspace{0.03cm}a}_{\hspace{0.02cm}{\bf k}}$ are the so-called normal field variables describing the soft bosonic degree of freedom of the system, and $\mathcal{Q}^{\hspace{0.03cm}a}$ is a color charge of the hard particle, which is a function of time $t$. Throughout the text we use the temporal gauge ($A_0$-gauge). The complete effective amplitude $\mathscr{T}^{\hspace{0.03cm}(2)\hspace{0.03cm}a\,a_{1}\hspace{0.03cm} a_{2}}_{\; {\bf k}_{1},\, {\bf k}_{2}} = f^{\hspace{0.03cm}a\,a_{1}\hspace{0.02cm}a_{2}}\,
\mathscr{T}^{\hspace{0.03cm}(2)}_{\; {\bf k}_{1},\, {\bf k}_{2}}$ has the following structure:
\begin{equation}
	\mathscr{T}^{\hspace{0.03cm}(2)}_{\; {\bf k}_{1},\, {\bf k}_{2}}
	=
	T^{\,(2)}_{\,{\bf k}_{1},\, {\bf k}_{2}}
	+
	\frac{1}{2}\,\biggl(\frac{1}
	{\omega^{\hspace{0.03cm}l}_{\hspace{0.03cm}{\bf k}_{1}} - {\bf v}\cdot {\bf k}_{1}}
	+
	\frac{1}
	{\omega^{\hspace{0.03cm}l}_{\hspace{0.03cm}{\bf k}_{2}} - {\bf v}\cdot {\bf k}_{2}}\biggr)
	\hspace{0.03cm}{\upphi}^{\hspace{0.03cm}\ast}_{\hspace{0.03cm}{\bf k}_{1}}\hspace{0.03cm}
	{\upphi}^{\phantom{\ast}}_{\hspace{0.03cm}{\bf k}_{2}}
	\label{eq:2w}
	\vspace{-0.3cm}
\end{equation}
\begin{align}
	+\;i\,\Biggl[
	\Biggl(&\frac{1}
	{\omega^{\hspace{0.03cm}l}_{\hspace{0.03cm}{\bf k}_{1} - {\bf k}_{2}}\! - {\bf v}\cdot ({\bf k}_{1} - {\bf k}_{2})}
	+
	\frac{1}
	{\omega^{\hspace{0.03cm}l}_{\hspace{0.03cm}{\bf k}_{1} - {\bf k}_{2}} -\omega^{\hspace{0.03cm}l}_{\hspace{0.03cm}{\bf k}_{1}} + \omega^{\hspace{0.03cm}l}_{\hspace{0.03cm}{\bf k}_{2}}}		
	\Biggr)\hspace{0.03cm}
	{\mathcal V}^{\phantom{\ast}}_{\,{\bf k}_{1},\, {\bf k}_{2},\, {\bf k}_{1} - {\bf k}_{2}} 
	{\upphi}^{\hspace{0.03cm}{\ast}}_{\hspace{0.03cm}{\bf k}_{1} - {\bf k}_{2}}
	\notag\\[1.5ex]
	-\,
	\Biggl(&\frac{1}
	{\omega^{\hspace{0.03cm}l}_{\hspace{0.03cm}{\bf k}_{2} - {\bf k}_{1}}\! - {\bf v}\cdot ({\bf k}_{2} - {\bf k}_{1})}
	+
	\frac{1}
	{\omega^{\hspace{0.03cm}l}_{\hspace{0.03cm}{\bf k}_{2} - {\bf k}_{1}} - \omega^{\hspace{0.03cm}l}_{\hspace{0.03cm}{\bf k}_{2}} + \omega^{\,l}_{\hspace{0.03cm}{\bf k}_{1}}}		
	\Biggr)\hspace{0.03cm}
	{\mathcal V}^{\,{\ast}}_{\,{\bf k}_{2},\, {\bf k}_{1},\, {\bf k}_{2} - {\bf k}_{1}} 
	{\upphi}^{\phantom{\ast}}_{\hspace{0.03cm}{\bf k}_{2} - {\bf k}_{1}}
	\Biggr].
	\notag
\end{align}
Here, $f^{\hspace{0.03cm}a\hspace{0.02cm}b\hspace{0.03cm}c}$ are the totally antisymmetric structure constants of the color Lie algebra $\mathfrak{su}(N_{c})$, $a,b,c = 1,\ldots,N_{c}$; 
$\omega^{\hspace{0.03cm}l}_{\hspace{0.03cm}{\bf k}}$ is the dispersion relation of the longitudinal mode of the collective excitations of QGP; ${\bf v}$ is velocity of the hard particle, which we consider to be fixed. 
The functions ${\upphi}^{\hspace{0.01cm}\ast}_{\hspace{0.02cm}{\bf k}}$ and ${\upphi}^{\phantom{\ast}}_{\hspace{0.02cm}{\bf k}}$ in the amplitude (\ref{eq:2w}) are elementary interaction vertices of the incoming and outgoing wave lines (plasmons) with a hard test color-charged particle G. An explicit form of the vertex function ${\upphi}^{\phantom{\ast}}_{\,{\bf k}}$ is
\begin{equation}
{\upphi}^{\phantom{\ast}}_{\hspace{0.03cm}{\bf k}}
=
g
\left(\frac{Z_{l}({\bf k})}
{2\hspace{0.03cm}\omega^{\hspace{0.03cm}l}_{\hspace{0.03cm}
		{\bf k}}}\right)^{\!\!1/2}\!\!\!
(v\cdot\epsilon^{\hspace{0.03cm}l}({\bf k})),
\label{eq:2ww}
\end{equation}
where the factor $Z_{l}({\bf k})$ is the residue of the effective gluon propagator at the longitudinal mode pole, $\epsilon^{\hspace{0.03cm}l}_{\mu}({\bf k}) = (\epsilon^{\hspace{0.03cm}l}_{0}({\bf k}),	\pmb{\epsilon}^{\hspace{0.03cm}l}({\bf k}))$ is the polarization vector in $A_{0}$\hspace{0.03cm}-\hspace{0.03cm}gauge (see Eq.\,(\ref{eq:7r})), and $v^{\hspace{0.02cm}\mu} = (1,{\bf v})$. An explicit form of the effective three-plasmon vertex function ${\mathcal V}^{\phantom{\ast}}_{\,{\bf k}_{1},\, {\bf k}_{2},\, {\bf k}_{3}}$ is presented in Appendix.\\
\indent The variables $c^{\ast\ \!\!a}_{\hspace{0.02cm}{\bf k}}, c^{\hspace{0.03cm}a}_{\hspace{0.02cm}{\bf k}}$ and $\mathcal{Q}^{\,a}$ obey the corresponding Hamilton equations: 
\begin{equation}
	\frac{\partial \hspace{0.02cm}c^{\hspace{0.03cm}\ast\,a}_{\hspace{0.02cm}{\bf k}}}{\partial\hspace{0.03cm} t}
	=
	-\hspace{0.03cm}i\hspace{0.05cm}\Bigl\{c^{\hspace{0.03cm}\ast\,a}_{\hspace{0.02cm}{\bf k}}\hspace{0.03cm},\hspace{0.03cm} {\mathcal H}^{(0)\!} + {\mathcal H}^{(4)}_{g\hspace{0.02cm}G\hspace{0.02cm}\rightarrow\hspace{0.02cm} g\hspace{0.02cm}G}\Bigr\},
	\quad
	\frac{\partial \hspace{0.02cm}c^{\,a^{\prime}}_{\hspace{0.02cm}{\bf k}^{\hspace{0.02cm}\prime}}}{\partial\hspace{0.03cm} t}
	=
	-\hspace{0.03cm}i\hspace{0.05cm}\Bigl\{c^{\phantom{\hspace{0.03cm}\ast} \!\!a^{\prime}}_{\hspace{0.02cm}{\bf k}^{\hspace{0.02cm}\prime}}\hspace{0.03cm},\hspace{0.03cm} {\mathcal H}^{(0)\!} + {\mathcal H}^{(4)}_{g\hspace{0.02cm}G\hspace{0.02cm}\rightarrow\hspace{0.02cm} g\hspace{0.02cm}G}\Bigr\},
	\label{eq:2e}
\end{equation}
\[
\frac{d \hspace{0.01cm}\mathcal{Q}^{\,a}}{d\hspace{0.03cm} t}
=
-\hspace{0.03cm}i\hspace{0.05cm}
\Bigl\{\mathcal{Q}^{\,a}\hspace{0.03cm},\hspace{0.03cm} {\mathcal H}^{(0)\!} + {\mathcal H}^{(4)}_{g\hspace{0.02cm}G\hspace{0.02cm}\rightarrow\hspace{0.02cm} g\hspace{0.02cm}G}\Bigr\},
\quad
\mathcal{Q}^{\,a}(t)|_{\hspace{0.02cm}t\hspace{0.03cm}=\hspace{0.03cm}t_{0}} = {\mathcal{Q}^{\,a}_{0}},
\]
where
\begin{equation}
	{\mathcal H}^{(0)} =  
	\!\int\!d\hspace{0.02cm}{\bf k}\, (\omega^{\hspace{0.03cm}l}_{\hspace{0.03cm}{\bf k}} - {\mathbf v}\cdot {\mathbf k})\ \!
	c^{\ast\hspace{0.03cm}a}_{\hspace{0.02cm}{\bf k}}\hspace{0.03cm} c^{\!\!\phantom{\ast} a}_{\hspace{0.03cm}{\bf k}}
	\label{eq:2r}
\end{equation}
is the free field Hamiltonian, braces denote the Lie-Poisson bracket 
\begin{equation}
	\vspace{0.03cm}
	\bigl\{F,\,G\bigr\} 
	=
	\int\! d\hspace{0.02cm}{\bf k\hspace{0.01cm}}'\!\hspace{0.02cm}
	\left\{\frac{\delta\hspace{0.01cm} F}{\delta\hspace{0.01cm} c^{\phantom{\ast}\!\!a}_{{\bf k}'}}
	\hspace{0.03cm}\frac{\delta\hspace{0.01cm}  G}{\delta\hspace{0.01cm} c^{\ast\ \!\!a}_{{\bf k}'}}
	\,-\,
	\frac{\delta\hspace{0.01cm}F}{\delta\hspace{0.01cm} 
		c^{\ast\ \!\!a}_{{\bf k}'}}\hspace{0.03cm}
	\frac{\delta\hspace{0.01cm}G}{\delta\hspace{0.01cm} c^{\phantom{\ast}\!\!a}_{{\bf k}'}}\right\}
	\,+\,
	i\,\frac{\partial F}{\,\partial\hspace{0.03cm} \mathcal{Q}^{\,a}}\hspace{0.03cm}
	\frac{\partial\hspace{0.03cm}G}{\,\partial\hspace{0.03cm} \mathcal{Q}^{\hspace{0.03cm}b}}
	\,f^{\hspace{0.03cm}a\hspace{0.03cm}b\hspace{0.03cm}c}\hspace{0.03cm}
	\mathcal{Q}^{\hspace{0.03cm}c}.
	\label{eq:2t}
	\vspace{0.03cm}
\end{equation}
The first term on the right-hand side is a standard canonical bracket.\\ 
\indent Fig.\,\ref{fig1} gives a diagrammatic interpretation of the different terms in the effective amplitude (\ref{eq:2w}).
The second and third diagrams represent the usual Compton scattering of soft bosonic excitations off a hard test particle induced by the second term on the right-hand side of the expression (\ref{eq:2w}). The incoming and outgoing wave lines in Fig.\,\ref{fig1} correspond to the normal field variables $c^{\hspace{0.03cm}a_{1}}_{\hspace{0.03cm}{\bf k}_{1}}$ and $c^{\ast\ \!\! a_{2}}_{\hspace{0.03cm}{\bf k}_{2}}$, respectively, and the horizontal double line between two interaction vertices corresponds to the ``propagator'' of the hard particle
\[
1/(\omega^{\hspace{0.02cm}l}_{\hspace{0.03cm}{\bf k}_{1}} - {\bf v}\cdot {\bf k}_{1}),
\]
which enters (\ref{eq:2w}) in the symmetrized form. The interaction vertices correspond to the functions ${\upphi}^{\hspace{0.02cm}\ast}_{\,{\bf k}_{1}}$ or
${\upphi}^{\phantom{\ast}}_{\,{\bf k}_{2}}$. The last graph in Fig.\,\ref{fig1} is related to the interaction of the hard particle with plasmons through the three-plasmon vertex function
${\mathcal V}^{\;a\,a_{1}\hspace{0.03cm}a_{2}}_{\;{\bf k},\, {\bf k}_{1},\, {\bf k}_{2}}$ with intermediate ``virtual'' oscillation to which corresponds  the factor in (\ref{eq:2w})
\[
\frac{1}
{\omega^{\hspace{0.03cm}l}_{\hspace{0.03cm}{\bf k}_{1} - {\bf k}_{2}}\! - {\bf v}\cdot ({\bf k}_{1} - {\bf k}_{2})}
\,+\,
\frac{1}
{\omega^{\hspace{0.03cm}l}_{\hspace{0.03cm}{\bf k}_{1} - {\bf k}_{2}} -\omega^{\hspace{0.03cm}l}_{\hspace{0.03cm}{\bf k}_{1}} + \omega^{\hspace{0.03cm}l}_{\hspace{0.03cm}{\bf k}_{2}}}.	
\]
This factor can also be written in a simpler form:
\[
1/(
\omega^{\hspace{0.03cm}l}_{\hspace{0.03cm}{\bf k}_{1} - {\bf k}_{2}} -\omega^{\hspace{0.03cm}l}_{\hspace{0.03cm}{\bf k}_{1}} + \omega^{\hspace{0.03cm}l}_{\hspace{0.03cm}{\bf k}_{2}}),
\]
if the so-called ``resonance frequency difference'' is exactly zero
\[
\Delta\hspace{0.02cm}\omega_{\hspace{0.03cm}{\mathbf k}_{1},\hspace{0.03cm}{\mathbf k}_{2}} 
\equiv
\omega^{\hspace{0.02cm}l}_{\hspace{0.03cm}{\bf k}_{1}} - \omega^{\hspace{0.02cm}l}_{\hspace{0.03cm}{\mathbf k}_{2}}
- {\mathbf v}\cdot (\hspace{0.03cm}{\mathbf k}_{1} - {\mathbf k}_{2})
= 0.
\]
\indent Finally, the first graph in Fig.\,\ref{fig1} corresponds to the first term  $T^{\hspace{0.03cm}(2)}_{\; {\bf k}_{1},\, {\bf k}_{2}}$ on the right-hand side of (\ref{eq:2w}), which is related to the process of direct interaction of two plasmons with a hard particle. In the physical system under consideration there is no double contact vertex function describing this scattering process, and therefore we should simply assume this term to be zero:
\[
T^{\hspace{0.03cm}(2)}_{\; {\bf k}_{1},\, {\bf k}_{2}} \equiv 0.
\] 
In conclusion of this section we also note that the Hamiltonian (\ref{eq:2q}) is a real function even without the fulfillment of the resonance condition 
$\Delta\hspace{0.02cm}\omega_{\hspace{0.03cm}{\mathbf k}_{1},\hspace{0.03cm}{\mathbf k}_{2}} = 0$ in the scattering processes that we are interested in.

\section{Classical scattering matrix}
\label{section_3}

This and next two sections are devoted to deriving  the classical scattering matrix for the scattering process of a hard color-charged particle off the soft bosonic QGP-excitations. Our further considerations will be largely based on the works of Zakhkarov and Schulman \cite{10.1007/3-540-11192-1_38, ZakShu85, Zakharov:1988}.\\ 
\indent In our case the following system of dynamical equations: 
\begin{align}
	&\frac{\partial \hspace{0.02cm}c^{\phantom{\hspace{0.03cm}\ast} \!\!a}_{\hspace{0.02cm}{\bf k}}}{\partial\hspace{0.03cm} t}
	=
	-\hspace{0.03cm}i\hspace{0.03cm}
	\bigl(\omega^{\hspace{0.02cm}l}_{\hspace{0.02cm}{\bf k}}
	- {\bf v}\cdot{\bf k}\bigr)
	\hspace{0.03cm}c^{\phantom{\hspace{0.03cm}\ast} \!\!a}_{\hspace{0.02cm}{\bf k}}
	-
	i\,\frac{\delta\hspace{0.03cm}{\mathcal H}_{int}}{\delta\hspace{0.01cm} c^{\ast\ \!\!a}_{\hspace{0.02cm}{\bf k}}}\hspace{0.03cm},
	\notag\\[1ex]	
	&\frac{\partial \hspace{0.02cm}c^{\hspace{0.03cm}\ast\,a}_{\hspace{0.02cm}{\bf k}}}{\partial\hspace{0.03cm} t}
	=
	i\hspace{0.03cm}
	\bigl(\omega^{\hspace{0.02cm}l}_{\hspace{0.02cm}{\bf k}}
	- {\bf v}\cdot{\bf k}\bigr)\hspace{0.03cm} 
	c^{\hspace{0.03cm}\ast\,a}_{\hspace{0.02cm}{\bf k}}
	+
	i\,\frac{\delta\hspace{0.03cm}{\mathcal H}_{int}}{\delta\hspace{0.01cm} c^{\ \!\!a}_{\hspace{0.02cm}{\bf k}}}\hspace{0.03cm}, 
	\label{eq:3q}\\[1ex]
	&\frac{d\hspace{0.02cm}\mathcal{Q}^{\hspace{0.03cm}a}}{d\hspace{0.02cm}t}
	=
	\frac{\!\partial\hspace{0.03cm}{\mathcal H}_{\hspace{0.03cm}int}}{\partial\hspace{0.03cm} \mathcal{Q}^{\hspace{0.03cm}b}}\,f^{\hspace{0.03cm}a\hspace{0.03cm}b\hspace{0.03cm}c}
	\hspace{0.03cm}\mathcal{Q}^{\hspace{0.03cm}c}
	\notag
\end{align}
are the starting ones in the construction of the classical scattering matrix. It is a consequence of the Hamilton equations (\ref{eq:2e}), of the definition of the free Hamiltonian (\ref{eq:2r}), and of the Lie-Poisson bracket (\ref{eq:2t}). Here,  ${\mathcal H}_{\hspace{0.03cm}int}$ is some interaction Hamiltonian. Following the reasoning \cite{10.1007/3-540-11192-1_38, ZakShu85, Zakharov:1988}, first we must introduce into consideration a system with an interaction, adiabatically switching off as $t\rightarrow \pm\infty$, i.e.,
\[
{\mathcal H} = {\mathcal H}_{0} + {\mathcal H}_{\hspace{0.03cm}int}\,{\rm e}^{-\epsilon|t|},
\quad \epsilon > 0.
\]
Solution of the equations (\ref{eq:3q}) turns asymptotically into the solution of the free-field equations:
\begin{equation}
	c^{\phantom{\hspace{0.03cm}\ast} \!\!a}_{\hspace{0.02cm}{\bf k}}(t) 
	\rightarrow
	c^{\pm\,a}_{\hspace{0.02cm}{\bf k}}(t)
	\equiv
	c^{\pm\,a}_{\hspace{0.02cm}{\bf k}}
	\ \!{\rm e}^{-i\hspace{0.03cm}(\omega^{\hspace{0.03cm}l}_{\hspace{0.02cm}{\bf k}}\hspace{0.03cm} -\hspace{0.03cm}{\bf v}\cdot{\bf k})\hspace{0.03cm}t},
	\qquad
	\mathcal{Q}^{\hspace{0.03cm}a}(t)\rightarrow\mathcal{Q}^{\pm\hspace{0.03cm}a},
	\label{eq:3w}
\end{equation}
where on the right-hand side the quantities $c^{\pm\,a}_{\hspace{0.02cm}{\bf k}}$ and $\mathcal{Q}^{\pm\hspace{0.03cm}a}$ are some constants. The functions $(c^{-\,a}_{\hspace{0.02cm}{\bf k}},
\,\mathcal{Q}^{-\hspace{0.03cm}a})$ and $(c^{+\,a}_{\hspace{0.02cm}{\bf k}},\,
\mathcal{Q}^{+\hspace{0.03cm}a})$ are not independent. There exists a nonlinear operator $\hat{S}_{\epsilon}$ relating the in- and out-fields and asymptotic color charges\footnote{\hspace{0.03cm}Sometimes we will use this convenient terminology commonly accepted in quantum field theory for the notation of asymptotic in- and out-field operators defined in the regions   $t\rightarrow -\infty$ and $t\rightarrow +\infty$, correspondingly (see, for example \cite{Schweber:1961zz}).  These operators, in particular, satisfy the free field commutation relations and equations.}. Here, the notations ``\hspace{0.03cm}in/out-'' are associated with the states that the signs ``$-/+$'' are assigned to. For further analysis we pass on to the so-called ``interaction representation''
\[
c^{\phantom{\hspace{0.03cm}\ast} \!\!a}_{\hspace{0.02cm}{\bf k}}(t)
=
\tilde{c}^{\phantom{\hspace{0.03cm}\ast} \!\!a}_{\hspace{0.02cm}{\bf k}}(t)
\ \!{\rm e}^{-i\hspace{0.03cm}(\omega^{\hspace{0.03cm}l}_{\hspace{0.02cm}{\bf k}}\hspace{0.03cm} -\hspace{0.03cm}{\bf v}\cdot{\bf k})\hspace{0.03cm}t},
\qquad
c^{\,\ast\,a}_{\hspace{0.02cm}{\bf k}}(t)
=
\tilde{c}^{\,\ast\,a}_{\hspace{0.02cm}{\bf k}}(t)
\ \!{\rm e}^{\,i\hspace{0.03cm}(\omega^{\hspace{0.03cm}l}_{\hspace{0.02cm}{\bf k}}\hspace{0.03cm} -\hspace{0.03cm}{\bf v}\cdot{\bf k})\hspace{0.03cm}t}.
\]
The equations of motion (\ref{eq:3q}) now take the form:
\begin{align}
	&\frac{\partial \hspace{0.02cm}\tilde{c}^{\phantom{\hspace{0.03cm}\ast} \!\!a}_{\hspace{0.02cm}{\bf k}}}{\partial\hspace{0.03cm} t}
	=
	-\,
	i\,\frac{\delta\hspace{0.03cm}\widetilde{\mathcal H}_{int}}{\delta\hspace{0.01cm} \tilde{c}^{\,\ast\ \!\!a}_{\hspace{0.02cm}{\bf k}}}\hspace{0.03cm}\,{\rm e}^{-\epsilon|t|},
	\notag\\[1ex]	
	&\frac{\partial \hspace{0.02cm}\tilde{c}^{\hspace{0.03cm}\ast\,a}_{\hspace{0.02cm}{\bf k}}}{\partial\hspace{0.03cm} t}
	=
	i\,\frac{\delta\hspace{0.03cm}\widetilde{\mathcal H}_{int}}{\delta\hspace{0.01cm} \tilde{c}^{\ \!\!a}_{\hspace{0.02cm}{\bf k}}}\,{\rm e}^{-\epsilon|t|}, 
	\notag\\[1ex]
	&\frac{d\hspace{0.02cm}\mathcal{Q}^{\hspace{0.03cm}a}}{d\hspace{0.02cm}t}
	=
	\frac{\!\partial\hspace{0.03cm}\widetilde{\mathcal H}_{\hspace{0.03cm}int}}{\partial\hspace{0.03cm} \mathcal{Q}^{\hspace{0.03cm}b}}\,f^{\hspace{0.03cm}a\hspace{0.03cm}b\hspace{0.03cm}c}
	\hspace{0.03cm}\mathcal{Q}^{\hspace{0.03cm}c}\,{\rm e}^{-\epsilon|t|},
	\notag	
\end{align}
where $\widetilde{\mathcal H}_{\hspace{0.03cm}int}$   is the interaction Hamiltonian expressed in terms of the new variables $\tilde{c}^{\phantom{\hspace{0.03cm}\ast} \!\!a}_{\hspace{0.02cm}{\bf k}}$ and $\tilde{c}^{\hspace{0.03cm}\ast\,a}_{\hspace{0.02cm}{\bf k}}$. These equations are equivalent to the integral ones governing the time evolution of the system under consideration:
\begin{equation} 
	\begin{split}
		&\tilde{c}^{\phantom{\hspace{0.03cm}\ast} \!\!a}_{\hspace{0.02cm}{\bf k}}(t)
		\,=\,
		c^{-\,a}_{\hspace{0.02cm}{\bf k}}
		\,-\,
		\frac{i}{2}\int\limits_{-\infty}^{t}\!d\tau\,
		\frac{\delta\hspace{0.03cm}\widetilde{\mathcal H}_{int}}{\delta\hspace{0.01cm} \tilde{c}^{\ast\ \!\!a}_{\hspace{0.02cm}{\bf k}}(\tau)}\hspace{0.03cm}\,{\rm e}^{-\epsilon|\tau|},\\
		&\tilde{c}^{\hspace{0.03cm}\ast\,a}_{\hspace{0.02cm}{\bf k}}(t)
		=
		(c^{-\,a}_{\hspace{0.02cm}{\bf k}})^{\ast}
		+
		\frac{i}{2}\int\limits_{-\infty}^{t}\!d\tau\,
		\frac{\delta\hspace{0.03cm}\widetilde{\mathcal H}_{int}}{\delta\hspace{0.01cm} \tilde{c}^{\phantom{\ast}\!\! a}_{\hspace{0.02cm}{\bf k}}(\tau)}\hspace{0.03cm}\,{\rm e}^{-\epsilon|\tau|},\\	
		&\mathcal{Q}^{\hspace{0.03cm}a}(t) 
		=
		\mathcal{Q}^{-\hspace{0.03cm}a}
		+
		\frac{1}{2}\int\limits_{-\infty}^{t}\!d\tau\,
		\frac{\!\partial\hspace{0.03cm}\widetilde{\mathcal H}_{\hspace{0.03cm}int}}{\partial\hspace{0.03cm} \mathcal{Q}^{\hspace{0.03cm}b}(\tau)}\,f^{\hspace{0.03cm}a\hspace{0.03cm}b\hspace{0.03cm}c}
		\hspace{0.03cm}\mathcal{Q}^{\hspace{0.03cm}c}(\tau)\,{\rm e}^{-\epsilon|\tau|}.
	\end{split}
	\label{eq:3e}
\end{equation}
Solutions of these integral equations can be formally represented in the following form:
\begin{equation} 
\begin{split}
		&\tilde{c}^{\phantom{\hspace{0.03cm}\ast} \!\!a}_{\hspace{0.02cm}
			{\bf k}}(t)
		=
		S_{\epsilon}(-\infty,t)[\hspace{0.03cm}c^{-\,a}_{\hspace{0.02cm}
			{\bf k}},(c^{-\,a}_{\hspace{0.02cm}{\bf k}})^{\ast},\mathcal{Q}^{-\hspace{0.03cm}a}],\\[1ex]
		&\tilde{c}^{\hspace{0.03cm}\ast\,a}_{\hspace{0.02cm}{\bf k}}(t)
		=
		S^{\,\ast}_{\epsilon}(-\infty,t)[\hspace{0.03cm}
		c^{-\,a}_{\hspace{0.02cm}{\bf k}},(c^{-\,a}_{\hspace{0.02cm}
			{\bf k}})^{\ast},\mathcal{Q}^{-\hspace{0.03cm}a}],\\[1ex]
		&\mathcal{Q}^{\hspace{0.03cm}a}(t) 
		=
		S_{\epsilon}(-\infty,t)[\hspace{0.03cm}
		c^{-\,a}_{\hspace{0.02cm}{\bf k}},(c^{-\,a}_{\hspace{0.02cm}
			{\bf k}})^{\ast},\mathcal{Q}^{-\hspace{0.03cm}a}].
\end{split}
\label{eq:3r}
\end{equation}
Here, in the square brackets, we indicate the asymptotic in-states that are mapped by the operator $S_{\varepsilon}(-\infty,t)$ into the time-dependent new variables $\tilde{c}^{\,\ast\,a}_{{\bf k}}(t), \tilde{c}^{\phantom{\ast} \!\!a}_{{\bf k}}(t)$ and $\mathcal{Q}^{\,a}(t)$. In the following, to avoid introducing new notations, the integral operators on the right-hand sides for the solutions $\tilde{c}^{\phantom{\ast} \!\!a}_{
{\bf k}}(t)$ and $\mathcal{Q}^{a}(t)$ are written by means of the same symbol $S_{\epsilon}(-\infty,t)[\,\ldots\,]$, although this is not quite correct.\\
\indent At finite $\epsilon$ and sufficiently small $c^{-\,a}_{\hspace{0.02cm}{\bf k}}$ and $\mathcal{Q}^{-\hspace{0.03cm}a}$, the integral operator $S_{\epsilon}(-\infty,t)$ can be obtained in the form of convergent series by the iteration of the integral equations (\ref{eq:3e}). In the work \cite{Zakharov:1988} the series obtained for the operator $S_{\epsilon}(-\infty,t)$ as $\epsilon\rightarrow +0$ was called the {\it classical transition matrix}. The limit $\epsilon\rightarrow +0$ was defined  for each term of the series. Such derived  expression is finite in the sense of generalized functions.\\
\indent Further letting $t\rightarrow +\infty$, one finds from (\ref{eq:3r})
\begin{align} 
		&c^{+\,a}_{\hspace{0.02cm}{\bf k}}
		=
		S_{\epsilon}[\hspace{0.03cm}c^{-\,a}_{\hspace{0.02cm}{\bf k}},(c^{-\,a}_{\hspace{0.02cm}{\bf k}})^{\ast},\mathcal{Q}^{-\hspace{0.03cm}a}],\notag\\[1ex]
		(&c^{+\,a}_{\hspace{0.02cm}{\bf k}})^{\ast}
		=
		S^{\,\ast}_{\epsilon}[\hspace{0.03cm}c^{-\,a}_{\hspace{0.02cm}{\bf k}},(c^{-\,a}_{\hspace{0.02cm}{\bf k}})^{\ast},\mathcal{Q}^{-\hspace{0.03cm}a}],	\label{eq:3t}\\[1ex]
		&\mathcal{Q}^{+\hspace{0.03cm}a}
		=
		S_{\epsilon}[\hspace{0.03cm}c^{-\,a}_{\hspace{0.02cm}{\bf k}},(c^{-\,a}_{\hspace{0.02cm}{\bf k}})^{\ast},\mathcal{Q}^{-\hspace{0.03cm}a}],\notag
\end{align}
where $S_{\epsilon}\equiv S_{\epsilon}(-\infty,+\infty)$. The corresponding limit of integral operator $S_{\epsilon}$ as $\epsilon\rightarrow +0$
\[
\mathcal{S} = \lim\limits_{\epsilon\rightarrow +0}S_{\epsilon}(-\infty,+\infty)
\]
was referred to as the {\it classical scattering matrix}.\\
\indent To conclude this section, we note that since the interaction is adiabatically switched off as $t\rightarrow -\infty$, the system, generally, in the asymptotic past can be in a non-equilibrium state (and even in an essentially non-equilibrium one). As will be shown in Section \ref{section_9}, the energy losses of an energetic particle propagating in a hot QCD medium depend on whether the system as $t\rightarrow -\infty$ is in thermal equilibrium or in a highly excited state.

\section{Plasmon scattering off a hard color-charged particle}
\label{section_4}

\begin{figure}[hbtp]
	\begin{center}
		\includegraphics[width=0.8\textwidth]{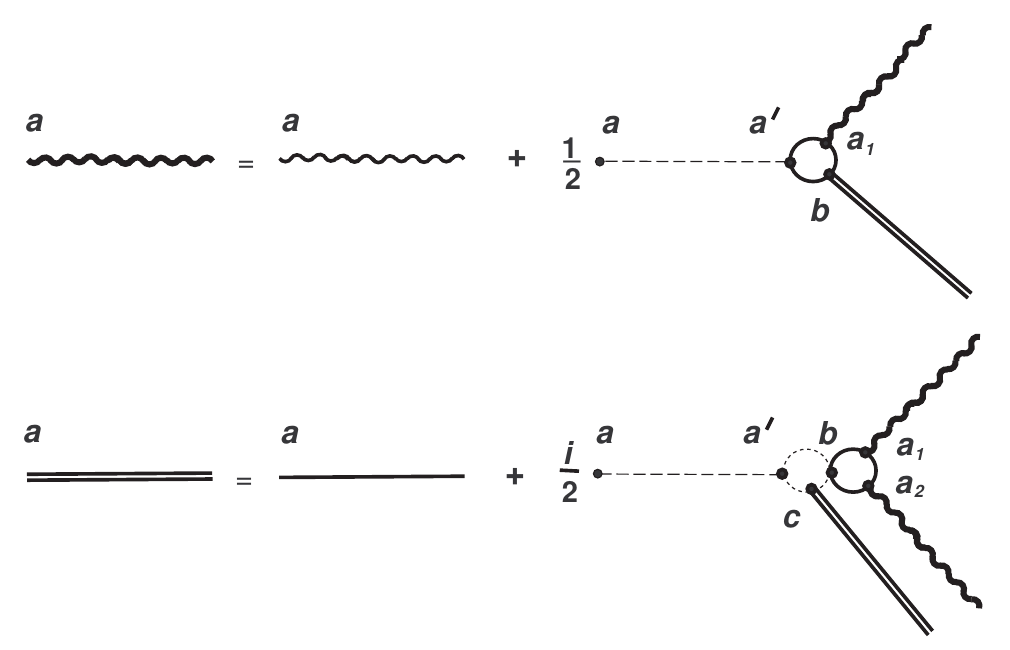}
	\end{center} 
	\vspace{-0.5cm}
	\caption{\small Graphical representation of two coupled integral equations (\ref{eq:4q}) and (\ref{eq:4w}) 
	}
	\label{fig2}
\end{figure}

Let us define the structure of the classical scattering matrix in the simplest case of the interaction Hamiltonian ${\mathcal H}_{\hspace{0.03cm}int} = {\mathcal  H}^{(4)}_{g\hspace{0.02cm}G\hspace{0.02cm}\rightarrow
	\hspace{0.02cm}g\hspace{0.02cm}G}$ that is quadratic in the field variables $\tilde{c}^{\phantom{\hspace{0.03cm}\ast} \!\!a}_{\hspace{0.02cm}{\bf k}}$ and $\tilde{c}^{\hspace{0.03cm}\ast\,a}_{\hspace{0.02cm}{\bf k}}$, and linear in the color charge $\mathcal{Q}^{\hspace{0.03cm}a}$, as it was defined by the expression (\ref{eq:2q}). In the interaction representation the first and  third integral equations in (\ref{eq:3e}) take the form
\begin{align} 
	&\tilde{c}^{\phantom{\hspace{0.03cm}\ast} \!\!a}_{\hspace{0.02cm}{\bf k}}(t)
	\,=\,
	c^{-\,a}_{\hspace{0.02cm}{\bf k}}
	+
	\frac{1}{2}\int\limits_{-\infty}^{t}\!d\tau\!
	\int\!d\hspace{0.02cm}{\bf k}_{1}\,
	\mathscr{T}^{\hspace{0.03cm}(2)\,b\,a\,a_{1}}_{\; {\bf k},\, {\bf k}_{1}}\,
	\tilde{c}^{\ \!a_{1}}_{{\bf k}_{1}}(\tau)
	\mathcal{Q}^{\,b}(\tau)\ \!{\rm e}^{i\hspace{0.03cm}	\Delta\hspace{0.02cm}\omega_{\hspace{0.03cm}{\mathbf k},\hspace{0.03cm}{\mathbf k}_{1}}\hspace{0.01cm}\tau \,-\, \epsilon\hspace{0.03cm}|\tau|},
	\label{eq:4q}\\[1ex]
	&\mathcal{Q}^{\hspace{0.03cm}a}(t) 
	=
	\mathcal{Q}^{-\hspace{0.03cm}a}
	\!+
	\frac{i}{2}\,f^{\hspace{0.03cm}a\hspace{0.03cm}b\hspace{0.03cm}c}\!
	\int\limits_{-\infty}^{t}\!d\tau\!
	\int\!d\hspace{0.02cm}{\bf k}_{1}\,d\hspace{0.02cm}{\bf k}_{2}\,
	\mathscr{T}^{\hspace{0.03cm}(2)\,b\,a_{1}\,a_{2}}_{\; {\bf k}_{1},\, {\bf k}_{2}}\,
	\tilde{c}^{\,\ast\ \!\!a_{1}}_{{\bf k}_{1}}(\tau)\,
	\tilde{c}^{\phantom{\hspace{0.03cm}\ast} \!\!a_{2}}_{\hspace{0.02cm}{\bf k}_{2}}(\tau)
	\mathcal{Q}^{\,c}(\tau)\ \!{\rm e}^{i\hspace{0.03cm}	\Delta\hspace{0.02cm}\omega_{\hspace{0.03cm}{\mathbf k}_{1},\hspace{0.03cm}{\mathbf k}_{2}}\hspace{0.01cm}\tau \,-\, \epsilon\hspace{0.03cm}|\tau|},
	\label{eq:4w}
\end{align}	
where the ``difference of resonant frequencies'' $\Delta\hspace{0.02cm}\omega_{\hspace{0.03cm}{\mathbf k},\hspace{0.03cm}{\mathbf k}_{1}}$ appears in the exponent of the integrands: 
\[
\Delta\hspace{0.02cm}\omega_{\hspace{0.03cm}{\mathbf k},\hspace{0.03cm}{\mathbf k}_{1}} 
\equiv
\omega^{\hspace{0.02cm}l}_{\hspace{0.03cm}{\bf k}_{1}} - \omega^{\hspace{0.02cm}l}_{\hspace{0.03cm}{\mathbf k}_{2}}
- {\mathbf v}\cdot (\hspace{0.03cm}{\mathbf k}_{1} - {\mathbf k}_{2}).
%
\]
Integral equations (\ref{eq:4q}) and (\ref{eq:4w}) can be symbolically represented in the graphical form as depicted in Fig.\,\ref{fig2}.
Explanations of the graphic elements are collected in Table\,\ref{table}.
\begin{table}[th]
	\centering%
	\begin{tabular}{|>{\centering\arraybackslash}b{5cm}|c|>{\centering
				\arraybackslash}b{3.5cm}|l|l|l|l|l|}
		\hline
		\textbf{\small{Name}} &\textbf{\small{Element of the diagram}} &\textbf{\small{Factor in the integral equations}}\\ \hline
		\vspace{0.03cm}
		\small{unknown\,normal\,field\,variable}
		&\includegraphics[width=0.2\textwidth]{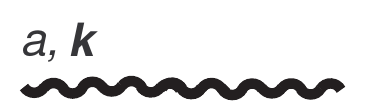} &$\tilde{c}^{\phantom{\hspace{0.03cm}\ast} \!\!a}_{\hspace{0.02cm}{\bf k}}(t)$\vspace{0.07cm}\\ \hline
		\small{unknow color charge}
		&\includegraphics[width=0.2\textwidth]{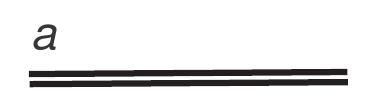}&$\mathcal{Q}^{\hspace{0.03cm}a}(t)$\vspace{0.04cm}\\ \hline
		\small{asymptotic field amplitude}
		&\includegraphics[width=0.2\textwidth]{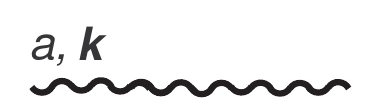}&$c^{-\,a}_{\hspace{0.02cm}{\bf k}}$\vspace{0.1cm}\\ \hline
		\small{asymptotic color charge}
		&\includegraphics[width=0.2\textwidth]{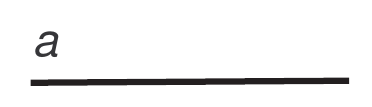}& $\mathcal{Q}^{-\hspace{0.03cm}a}$\vspace{0.1cm}\\ \hline
		\small{exponential factor}
		&\includegraphics[width=0.2\textwidth]{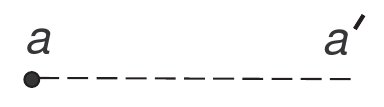}&
		$\delta^{\hspace{0.03cm}a\hspace{0.03cm}a^{\prime}}{\rm e}^{\,i\tau\Delta\hspace{0.02cm}\omega_{\hspace{0.03cm}{\mathbf k},\hspace{0.03cm}{\mathbf k}_{1}} - \epsilon|\tau|}$\vspace{0.04cm}\\ \hline
		\hspace*{0pt}\small{complete effective amplitude}
		\vspace{0.3cm}
		&\includegraphics[width=0.15\textwidth]{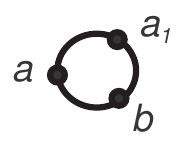}& $\mathscr{T}^{\hspace{0.03cm}(2)\,b\,a\,a_{1}}_{\; {\bf k},\, {\bf k}_{1}}$\vspace{0.5cm}\\ \hline
		\small{antisymmetric structure constants}\vspace{0.27cm}    
		&\includegraphics[width=0.15\textwidth]{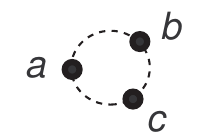}& $f^{\hspace{0.03cm}a\hspace{0.03cm}b\hspace{0.03cm}c}$\vspace{0.5cm}\\ \hline
	\end{tabular}
	\caption{\small Diagrammatic elements for graphical interpretation of integral equations (\ref{eq:4q}) and (\ref{eq:4w})}
\label{table}
\end{table}
\!\!\!The diagrammatic representation of a self-consistent system of two integral equations in general does not have any specific physical meaning, but rather serves as a simple graphical illustration. Such a representation is convenient because it provides an ability to attribute certain graphical diagram to each term of the series arising from iteration of the integral equations (\ref{eq:4q}) and (\ref{eq:4w}).\\ 
\indent The graphical representation of integral equation for unknown quantity (more precisely, for the normal field variable) goes back to the early classical works devoted to the statistical description of weakly turbulent wave fields (see, for example, Wyld \cite{Wyld:1961}, Zakharov and L’vov \cite{Zakharov:1975}, Zakharov and Schulman \cite{Zakharov:1988}, etc.). In our case, due to the consideration of the hard modes of excitations in the system, we have an additional integral equation. We attempted to give a graphical interpretation of the integral equations using the ideas of the works mentioned above with an appropriate generalization to the color degrees of freedom.\\
\indent For our purposes it is sufficient to define the first order iteration of Eq.\,(\ref{eq:4q}), i.e., on the right-hand side we just make the replacement: $\tilde{c}^{\phantom{\hspace{0.03cm}\ast} \!\!a}_{\hspace{0.02cm}{\bf k}}(\tau)\rightarrow c^{-\,a}_{\hspace{0.02cm}{\bf k}}$ and $\mathcal{Q}^{\hspace{0.03cm}a}(\tau)\rightarrow 
\mathcal{Q}^{-\hspace{0.03cm}a}$, then
\begin{equation}
	\tilde{c}^{\phantom{\hspace{0.03cm}\ast} \!\!a}_{\hspace{0.02cm}{\bf k}}(t)
	\,=\,
	c^{-\,a}_{\hspace{0.02cm}{\bf k}}
	\,+\,
	\frac{1}{2}
	\int\!d\hspace{0.02cm}{\bf k}_{1}
	\Biggl(\;\int\limits_{-\infty}^{t}\!d\tau\!\  {\rm e}^{\,i\hspace{0.03cm}\Delta\hspace{0.02cm}\omega_{\hspace{0.03cm}{\mathbf k},\hspace{0.03cm}{\mathbf k}_{1}}\hspace{0.01cm}\tau \,-\, \epsilon|\tau|}\Biggr)\hspace{0.03cm}
	\mathscr{T}^{\hspace{0.03cm}(2)\,b\,a\,a_{1}}_{\; {\bf k},\, {\bf k}_{1}}
	c^{-\,a_{1 }}_{\hspace{0.02cm}{\bf k}_{1}}
	\hspace{0.03cm}\mathcal{Q}^{-\hspace{0.03cm}b}.
	\label{eq:4r}
\end{equation}
Here, the time dependence is collected in a separate multiplier. Let us analyze the integral over~$\tau$. For definiteness, we assume that 
$t > 0$ and therefore 
\[
\int\limits_{-\infty}^{t}\!d\tau\!\  
{\rm e}^{\,i\hspace{0.03cm}\Delta\hspace{0.02cm}\omega_{\hspace{0.03cm}
		{\mathbf k},\hspace{0.03cm}{\mathbf k}_{1}}\hspace{0.01cm}\tau \,-\, \epsilon\hspace{0.03cm}|\tau|}
=
\int\limits_{-\infty}^{0}\!d\tau\!\  {\rm e}^{\,i\hspace{0.03cm}\Delta\hspace{0.02cm}\omega_{\hspace{0.03cm}{\mathbf k},\hspace{0.03cm}{\mathbf k}_{1}}\hspace{0.01cm}\tau \,+\, \epsilon\hspace{0.02cm}\tau}
+
\int\limits^{t}_{0}\!d\tau\!\  
{\rm e}^{\,i\hspace{0.03cm}\Delta\hspace{0.02cm}\omega_{\hspace{0.03cm}
		{\mathbf k},\hspace{0.03cm}{\mathbf k}_{1}}\hspace{0.01cm}\tau \,-\, \epsilon\hspace{0.02cm}\tau}	
\]
\begin{align} 
	&=
	\frac{1}{i\hspace{0.02cm}\Delta\hspace{0.02cm}\omega_{\hspace{0.03cm}{\mathbf k},\hspace{0.03cm}{\mathbf k}_{1}} + \epsilon}
	+
	\left(
	\frac{1}{i\hspace{0.02cm}\Delta\hspace{0.02cm}\omega_{\hspace{0.03cm}{\mathbf k},\hspace{0.03cm}{\mathbf k}_{1}} - \epsilon}\,
	{\rm e}^{\,(i\hspace{0.03cm}\Delta\hspace{0.02cm}\omega_{\hspace{0.03cm}
			{\mathbf k},\hspace{0.03cm}{\mathbf k}_{1}} \,-\, \epsilon)\hspace{0.03cm}t}
	\,-\,
	\frac{1}{i\hspace{0.02cm}\Delta\hspace{0.02cm}\omega_{\hspace{0.03cm}{\mathbf k},\hspace{0.03cm}{\mathbf k}_{1}} - \epsilon}\right)
	\notag\\[1ex]
	&= 
	\frac{2\hspace{0.03cm}\epsilon}{(\Delta\hspace{0.02cm}\omega_{\hspace{0.03cm}{\mathbf k},\hspace{0.03cm}{\mathbf k}_{1}})^{\hspace{0.03cm}2} + \epsilon^{\hspace{0.03cm}2}}
	\,+\,
	\frac{1}{i}\,\frac{1}{\Delta\hspace{0.02cm}\omega_{\hspace{0.03cm}{\mathbf k},\hspace{0.03cm}{\mathbf k}_{1}} + i\hspace{0.02cm}\epsilon}\,
	{\rm e}^{\,(i\hspace{0.03cm}\Delta\hspace{0.02cm}\omega_{\hspace{0.03cm}
			{\mathbf k},\hspace{0.03cm}{\mathbf k}_{1}} \,-\, \epsilon)\hspace{0.03cm}t}.	
	\notag
\end{align}
By using the following limits \cite{Vladimirov:1971}: 
\[
\lim\limits_{\epsilon\rightarrow +0}\frac{\epsilon}{x^{\hspace{0.03cm}2} + \epsilon^{\hspace{0.03cm}2}}
=
\pi\hspace{0.03cm}\delta(x), 
\qquad
\lim\limits_{t\rightarrow +\infty}\frac{{\rm e}^{\,i\hspace{0.03cm}x\hspace{0.03cm}t}}{x + i\hspace{0.03cm}\epsilon}
= 0,
\]
we find the required expression for the integral at hand
\[
\lim\limits_{t\rightarrow +\infty}\,\lim\limits_{\epsilon\rightarrow +0}
\int\limits_{-\infty}^{t}\!d\tau\!\  
{\rm e}^{\,i\hspace{0.03cm}\Delta\hspace{0.02cm}\omega_{\hspace{0.03cm}
		{\mathbf k},\hspace{0.03cm}{\mathbf k}_{1}}\hspace{0.01cm}\tau \,-\, \epsilon\hspace{0.03cm}|\tau|}
=
2\hspace{0.02cm}\pi\hspace{0.03cm}\delta(\Delta\hspace{0.02cm}\omega_{\hspace{0.03cm}{\mathbf k},\hspace{0.03cm}{\mathbf k}_{1}}).
\]
Thus letting, ${\epsilon\rightarrow +0}$ and ${t\rightarrow +\infty}$, one obtains from (\ref{eq:4r})
\begin{equation}
	c^{+\,a}_{\hspace{0.02cm}{\bf k}}
	\,=\,
	c^{-\,a}_{\hspace{0.02cm}{\bf k}}
	+
	\frac{1}{2}\int\!d\hspace{0.02cm}{\bf k}_{1}\,
	\mathscr{T}^{\hspace{0.03cm}(2)\,b\,a\,a_{1}}_{\; {\bf k},\, {\bf k}_{1}} c^{-\,a_{1}}_{\hspace{0.02cm}{\bf k}_{1}}\hspace{0.03cm}
	\mathcal{Q}^{-\hspace{0.03cm}b}\,
	2\hspace{0.02cm}\pi\hspace{0.03cm}\delta(\Delta\hspace{0.02cm}\omega_{\hspace{0.03cm}{\mathbf k},\hspace{0.03cm}{\mathbf k}_{1}})
	\equiv
	S[\hspace{0.03cm}c^{-\,a}_{\hspace{0.02cm}{\bf k}},(c^{-\,a}_{\hspace{0.02cm}{\bf k}})^{\ast},\mathcal{Q}^{-\hspace{0.03cm}a}].
	\label{eq:4t}
\end{equation}
This expression defines the classical scattering matrix in the first nontrivial approximation. Similar reasoning for the second integral equation (\ref{eq:4w}) leads us to the following relation in the same iteration order, which supplements (\ref{eq:4t}):
\begin{equation}
	\mathcal{Q}^{+\hspace{0.03cm}a}
	=
	\mathcal{Q}^{-\hspace{0.03cm}a}
	\!+
	\frac{i}{2}\,f^{\hspace{0.03cm}a\hspace{0.03cm}b\hspace{0.03cm}c}\!
	\int\!d\hspace{0.02cm}{\bf k}_{1}\,d\hspace{0.02cm}{\bf k}_{2}\,
	\mathscr{T}^{\hspace{0.03cm}(2)\,b\,a_{1}\,a_{2}}_{\; {\bf k}_{1},\, {\bf k}_{2}}\,
	(c^{-\,a_{1}}_{\hspace{0.02cm}{\bf k}_{1}})^{\ast}\,
	c^{-\,a_{2}}_{\hspace{0.02cm}{\bf k}_{2}}
	\mathcal{Q}^{-\hspace{0.03cm}c}\ \!	2\hspace{0.02cm}\pi\hspace{0.03cm}\delta(\Delta\hspace{0.02cm}\omega_{\hspace{0.03cm}{\mathbf k}_{1},\hspace{0.03cm}{\mathbf k}_{2}}).
	\label{eq:4y}
\end{equation}

\section{Explicit form of classical scattering matrix}
\label{section_5}

To determine the effective classical current it is necessary to know an explicit form of the classical scattering matrix, whereas in the expressions (\ref{eq:4t}) and (\ref{eq:4y}) it is given in the form of some integral operator. Let us try to define the explicit form of the classical scattering matrix in analogy to quantum field theory. As is well known, the relation between asymptotic states of any in- and out-field operators is given by the quantum field $S$-matrix \cite{Schweber:1961zz, Bogolyubov:1975ps, Bogolubov:1990}:
\[
\hat{\phi}^{\,out}(x) = \hat{S}^{\hspace{0.03cm}\dagger}\hat{\phi}^{\,in}(x)\hspace{0.03cm}\hat{S}.
\]  
Further, if we introduce the general form of quantum $S$-matrix to be an exponent of some phase function $\hat{T}$, to take into account its unitarity, (see, for example, \cite{Lehmann:1957})
\begin{equation}
	\hat{S} = {\rm e}^{\,i\hspace{0.03cm}\hat{T}},
	\label{eq:5q}
\end{equation}
where $\hat{T}$ is a hermitian operator, then the relation connecting the in- and out-field operators can be expanded in a series of commutators
\begin{equation}
	\hat{\phi}^{\,out}(x) = {\rm e}^{\,-i\hspace{0.03cm}\hat{T}}\hat{\phi}^{\,in}(x)\hspace{0.03cm}{\rm e}^{\,i\hspace{0.03cm}\hat{T}}
	\label{eq:5w}
\end{equation} 
\[
=
\hat{\phi}^{\,in}(x) + \frac{i}{1!}\,[\hspace{0.03cm}\hat{\phi}^{\,in},\hat{T}\hspace{0.03cm}]
\,+\,
\frac{i^{\hspace{0.03cm}2}}{\hspace{0.03cm}2\hspace{0.03cm}!}\,
[\hspace{0.03cm}[\hat{\phi}^{\,in},\hat{T}\hspace{0.03cm}],\hat{T}
\hspace{0.03cm}]
\,+\,
\frac{i^{\hspace{0.03cm}3}}{3\hspace{0.03cm}!}\,[\hspace{0.03cm}
[\hspace{0.03cm}[\hat{\phi}^{\,in},\hat{T}\hspace{0.03cm}],\hat{T}
\hspace{0.03cm}],\hat{T}\hspace{0.03cm}]
\,+\,\dots\,.
\]
By analogy with (\ref{eq:5q}), we will search for the classical $\mathcal{S}$-matrix in the form of an exponential function
\begin{equation}
	\mathcal{S} = {\rm e}^{\,i\hspace{0.03cm}\mathcal{T}},
	\label{eq:5e}
\end{equation}
where now $\mathcal{T} = \mathcal{T}^{\,\ast}$, 
and replace the quantum commutators in (\ref{eq:5w}) by the Lie-Poisson bracket: $[\cdot,\cdot]\rightarrow \{\cdot,\cdot\}$. The latter was defined by Eq.\,(\ref{eq:2t}). We write out the Lie-Poisson bracket in the new asymptotic variables\footnote{\hspace{0.03cm}The mappings (\ref{eq:3w}) are a formal canonical transformation, and in the new variables the {\it complete} Hamiltonian $\mathcal{H}$ has the form
	\[
	{\mathcal H} =  
	\!\int\!d\hspace{0.02cm}{\bf k}\, (\omega^{\hspace{0.03cm}l}_{\hspace{0.03cm}{\bf k}} - {\mathbf v}\cdot {\mathbf k})\ \!
	(c^{\pm\,a}_{\hspace{0.02cm}{\bf k}})^{\ast}
	c^{\pm\,a}_{\hspace{0.02cm}{\bf k}}.
	\]
}
$c^{-\,a}_{\hspace{0.02cm}{\bf k}}$, $(c^{-\,a}_{\hspace{0.02cm}{\bf k}})^{\ast}$ and  $\mathcal{Q}^{-\hspace{0.03cm}a}$:
\[
\bigl\{F,\,G\bigr\} 
=
\int\! d\hspace{0.02cm}{\bf k\hspace{0.01cm}}'\!\hspace{0.02cm}
\left\{\frac{\delta\hspace{0.01cm} F}{\delta\hspace{0.01cm} c^{\phantom{\ast}\!\!\!-\,c}_{{\bf k}'}}
\hspace{0.03cm}\frac{\delta\hspace{0.01cm} G}
{\delta\hspace{0.01cm}(c^{-\,c}_{\hspace{0.02cm}{\bf k}'})^{\ast}}
\,-\,
\frac{\delta\hspace{0.01cm}F}
{\delta\hspace{0.01cm}(c^{-\,c}_{\hspace{0.02cm}{\bf k}'})^{\ast}}
\hspace{0.03cm}
\frac{\delta\hspace{0.01cm}G}{\delta\hspace{0.01cm} c^{\phantom{\ast}\!\!\!-\,c}_{{\bf k}'}}\right\}
\,+\,
i\,\frac{\partial F}{\,\partial\hspace{0.03cm}\mathcal{Q}^{-\hspace{0.03cm}a}}\hspace{0.03cm}
\frac{\partial\hspace{0.03cm}G}{\,\partial\hspace{0.03cm} \mathcal{Q}^{-\hspace{0.03cm}b}}
\,f^{\hspace{0.03cm}a\hspace{0.03cm}b\hspace{0.03cm}c}\hspace{0.03cm}
\mathcal{Q}^{-\hspace{0.03cm}c}.
\]
Then the right-hand side of the first and the last relations in (\ref{eq:3t}) in the limit ${\epsilon\rightarrow +0}$ can be formally represented as the following series
\begin{align}
	&c^{+\,a}_{\hspace{0.02cm}{\bf k}}
	\,=\,
	c^{-\,a}_{\hspace{0.02cm}{\bf k}}
	+
	\frac{i}{1!}\,\{\hspace{0.03cm}c^{-\,a}_{\hspace{0.02cm}{\bf k}},\mathcal{T}\hspace{0.03cm}\}
	\,+\,
	\frac{i^{\hspace{0.02cm}2}}{\hspace{0.03cm}2\hspace{0.03cm}!}\,
	\{\hspace{0.03cm}\{c^{-\,a}_{\hspace{0.02cm}
		{\bf k}},\mathcal{T}\hspace{0.03cm}\},\mathcal{T}
	\hspace{0.03cm}\}
	\,+\,
	\frac{i^{\hspace{0.02cm}3}}{3\hspace{0.03cm}!}\,\{\hspace{0.03cm}
	\{\hspace{0.03cm}\{c^{-\,a}_{\hspace{0.02cm}{\bf k}},\mathcal{T}\hspace{0.03cm}\},\mathcal{T}
	\hspace{0.03cm}\},\mathcal{T}\hspace{0.03cm}\}
	\,+\,\dots\,,
	\label{eq:5r}\\[1ex]
	&\mathcal{Q}^{+\hspace{0.03cm}a} 
	=
	\mathcal{Q}^{-\hspace{0.03cm}a}
	+
	\frac{i}{1!}\,\{\hspace{0.03cm}\mathcal{Q}^{-\hspace{0.03cm}a},
	\mathcal{T}\hspace{0.03cm}\}
	\,+\,
	\frac{i^{\hspace{0.02cm}2}}{\hspace{0.03cm}2\hspace{0.03cm}!}\,
	\{\hspace{0.03cm}\{\mathcal{Q}^{-\hspace{0.03cm}a},\mathcal{T}
	\hspace{0.03cm}\},\mathcal{T}
	\hspace{0.03cm}\}
	\,+\,
	\frac{i^{\hspace{0.02cm}3}}{3\hspace{0.03cm}!}\,\{\hspace{0.03cm}
	\{\hspace{0.03cm}\{\mathcal{Q}^{-\hspace{0.03cm}a},\mathcal{T}
	\hspace{0.03cm}\},\mathcal{T}
	\hspace{0.03cm}\},\mathcal{T}\hspace{0.03cm}\}
	\,+\,\dots\,.
	\label{eq:5t}
\end{align}
These series actually represent some canonical transformation. Discussions of such transformations in the case of analytical mechanics can be found in textbooks \cite{Sudarshan:1974, Medvedev:2007}. They are closely related to one-parameter subgroup of general canonical transformations, in which the function $\mathcal{T}$ (in our case a functional) plays the role of {\it generator} of the subgroup. In particular, in the work \cite{Kim:2024} the classical quantity $\mathcal{T}$ was called ``scattering generator''. However, the examples considered in \cite{Sudarshan:1974, Medvedev:2007} assume that $\mathcal{T}$ is a function with a fixed functional form. In our case, the functional $\mathcal{T}$ itself is an unknown quantity subject to determination.\\ 
\indent Let us seek the function $\mathcal{T}$ in the form of the most general integro-power series expansion in the normal in-field  variables $c^{-\,a}_{\hspace{0.02cm}{\bf k}},\,(c^{-\,a}_{\hspace{0.02cm}{\bf k}})^{\ast}$ and in the asymptotic color charge ${\mathcal Q}^{\,-a}$
\begin{equation}
\mathcal{T} = 
{F}^{\,a}\hspace{0.02cm}
{\mathcal Q}^{-\hspace{0.02cm}a}
\label{eq:5y}
\vspace{-0.3cm}
\end{equation}	
\begin{align*}
\,+
\!&\int\!d\hspace{0.02cm}{\bf k}_{1}
\big[\hspace{0.04cm}g^{\,a_{1}}_{\, {\mathbf k}_{1}}c^{-\,a_{1}}_{\hspace{0.02cm}{\bf k}_{1}}
+
g^{\hspace{0.03cm}\ast\hspace{0.03cm}a_{1}}_{\, {\mathbf k}_{1}}(c^{-\,a_{1}}_{\hspace{0.02cm}{\bf k}_{1}})^{\ast}\bigr]
+
\!\int\!d\hspace{0.02cm}{\bf k}_{1}
\big[\hspace{0.04cm}f^{\,a_{1}\hspace{0.02cm}b}_{\, 
	{\mathbf k}_{1}}c^{-\,a_{1}}_{\hspace{0.02cm}{\bf k}_{1}}
+
f^{\hspace{0.03cm}\ast\hspace{0.03cm}a_{1}\hspace{0.02cm}b}_{\, 
	{\mathbf k}_{1}}(c^{-\,a_{1}}_{\hspace{0.02cm}{\bf k}_{1}})^{\ast}\bigr]
{\mathcal Q}^{-\hspace{0.02cm}b}\\[1ex]
+ &\!\int\!d\hspace{0.02cm}{\bf k}_{1}\hspace{0.02cm} d\hspace{0.02cm}{\bf k}_{2}\! 
\left[\hspace{0.04cm}
g^{\,(1)\,a_{1}a_{2}}_{\,{\mathbf k}_{1},\,{\mathbf k}_{2}}
c^{-\,a_{1}}_{\hspace{0.02cm}{\bf k}_{1}}
c^{-\,a_{2}}_{\hspace{0.02cm}{\bf k}_{2}}
+
g^{\,(2)\,a_{1}a_{2}}_{\,{\mathbf k}_{1},\,{\mathbf k}_{2}}
(c^{-\,a_{1}}_{\hspace{0.02cm}{\bf k}_{1}})^{\ast}
c^{-\,a_{2}}_{\hspace{0.02cm}{\bf k}_{2}}
+
g^{\hspace{0.03cm}\ast\,(1)\,a_{1}a_{2}}_{\,{\mathbf k}_{1},\,{\mathbf k}_{2}}
(c^{-\,a_{1}}_{\hspace{0.02cm}{\bf k}_{1}})^{\ast}
(c^{-\,a_{2}}_{\hspace{0.02cm}{\bf k}_{2}})^{\ast}\right] 
\end{align*}
\vspace{-0.3cm}
\[
+ \!\int\!d\hspace{0.02cm}{\bf k}_{1}\hspace{0.02cm} d\hspace{0.02cm}{\bf k}_{2} 
\left[\hspace{0.04cm}
G^{\,(1)\,a_{1}a_{2}\hspace{0.02cm}b}_{\,{\mathbf k}_{1},\,{\mathbf k}_{2}}
c^{-\,a_{1}}_{\hspace{0.02cm}{\bf k}_{1}}
c^{-\,a_{2}}_{\hspace{0.02cm}{\bf k}_{2}}
+
G^{\,(2)\,a_{1}a_{2}\hspace{0.02cm}b}_{\,{\mathbf k}_{1},\,{\mathbf k}_{2}}
(c^{-\,a_{1}}_{\hspace{0.02cm}{\bf k}_{1}})^{\ast}
c^{-\,a_{2}}_{\hspace{0.02cm}{\bf k}_{2}}
\right.
\]
\[
\left.
+\,
G^{\hspace{0.03cm}\ast\,(1)\,a_{1}a_{2}\hspace{0.02cm}b}_{\,{\mathbf k}_{1},\,{\mathbf k}_{2}}
(c^{-\,a_{1}}_{\hspace{0.02cm}{\bf k}_{1}})^{\ast}
(c^{-\,a_{2}}_{\hspace{0.02cm}{\bf k}_{2}})^{\ast}\right] 
{\mathcal Q}^{-\hspace{0.02cm}b}
+\,\ldots\,.
\]
Within accepted approximation it is sufficient to consider only the second term on the right-hand sides of (\ref{eq:5r}) and (\ref{eq:5t}),
then we have, respectively, 
\begin{align}
\{\hspace{0.03cm}&c^{-\,a}_{\hspace{0.02cm}{\bf k}},\mathcal{T}\hspace{0.03cm}\}
=
\frac{\delta\hspace{0.03cm}\mathcal{T}}{\delta\hspace{0.02cm} (c^{-\,a}_{\hspace{0.02cm}{\bf k}})^{\ast}}
=
g^{\,\ast\,a}_{\,{\mathbf k}}
+
f^{\hspace{0.03cm}\ast\hspace{0.03cm}a\hspace{0.02cm}b}_{\, 
	{\mathbf k}}\hspace{0.03cm}{\mathcal Q}^{-\hspace{0.02cm}b}
\notag\\[1ex]
+ 
&\int\!d\hspace{0.02cm}{\bf k}_{1}\! 
\left[\hspace{0.04cm}
g^{\,(2)\,a\,a_{1}}_{\,{\mathbf k},\,{\mathbf k}_{1}}
c^{-\,a_{1}}_{\hspace{0.02cm}{\bf k}_{1}}
+
2\hspace{0.03cm}g^{\hspace{0.03cm}\ast\,(1)\,a\,a_{1}}_{\,{\mathbf k},\,{\mathbf k}_{1}}
(c^{-\,a_{1}}_{\hspace{0.02cm}{\bf k}_{1}})^{\ast}
\right]\notag\\[1ex] 
+ 
&\int\!d\hspace{0.02cm}{\bf k}_{1}\! 
\left[\hspace{0.04cm}
G^{\,(2)\,a\,a_{1}\hspace{0.02cm}b}_{\,{\mathbf k},\,{\mathbf k}_{1}}
c^{-\,a_{1}}_{\hspace{0.02cm}{\bf k}_{1}}
+
2\hspace{0.03cm}G^{\hspace{0.03cm}\ast\,(1)\,a\,a_{1}\hspace{0.02cm}b}_{\,
	{\mathbf k},\,{\mathbf k}_{1}}
(c^{-\,a_{1}}_{\hspace{0.02cm}{\bf k}_{1}})^{\ast}
\right]\!{\mathcal Q}^{-\hspace{0.02cm}b}
+\,\ldots
\notag
\vspace{0.3cm}	
\end{align}
and
\vspace{-0.3cm}
\[
\{\hspace{0.03cm}\mathcal{Q}^{-\hspace{0.03cm}a},\mathcal{T}\hspace{0.03cm}\}
=
\frac{\!\partial\hspace{0.03cm}\mathcal{T}}{\partial\hspace{0.03cm} \mathcal{Q}^{-\hspace{0.03cm}b}}\,f^{\hspace{0.03cm}a\hspace{0.03cm}b
	\hspace{0.03cm}c}\hspace{0.03cm}\mathcal{Q}^{-\hspace{0.03cm}c}
=
f^{\hspace{0.03cm}a\hspace{0.03cm}b	\hspace{0.03cm}c}\hspace{0.03cm}
{F}^{\,b}\hspace{0.02cm}{\mathcal Q}^{-\hspace{0.02cm}c}
+\,
f^{\hspace{0.03cm}a\hspace{0.03cm}b\hspace{0.03cm}c}
\!\int\!d\hspace{0.02cm}{\bf k}_{1}
\big[\hspace{0.04cm}f^{\,a_{1}\hspace{0.03cm}b}_{\, 
	{\mathbf k}_{1}}c^{-\,a_{1}}_{\hspace{0.02cm}{\bf k}_{1}}
+
f^{\hspace{0.03cm}\ast\hspace{0.03cm}a_{1}\hspace{0.03cm}b}_{\, 
	{\mathbf k}_{1}}(c^{-\,a_{1}}_{\hspace{0.02cm}{\bf k}_{1}})^{\ast}\bigr]
{\mathcal Q}^{-\hspace{0.02cm}c}
\]
\[
+\, 
f^{\hspace{0.03cm}a\hspace{0.03cm}b\hspace{0.03cm}c}\!\!
\int\!d\hspace{0.02cm}{\bf k}_{1}\hspace{0.02cm} d\hspace{0.02cm}{\bf k}_{2}\! 
\left[\hspace{0.04cm}
G^{\,(1)\,a_{1}a_{2}\hspace{0.02cm}b}_{\,{\mathbf k}_{1},\,{\mathbf k}_{2}}
c^{-\,a_{1}}_{\hspace{0.02cm}{\bf k}_{1}}
c^{-\,a_{2}}_{\hspace{0.02cm}{\bf k}_{2}}
\!+
G^{\,(2)\,a_{1}a_{2}\hspace{0.02cm}b}_{\,{\mathbf k}_{1},\,{\mathbf k}_{2}}
(c^{-\,a_{1}}_{\hspace{0.02cm}{\bf k}_{1}})^{\ast}
c^{-\,a_{2}}_{\hspace{0.02cm}{\bf k}_{2}}
\!+
G^{\hspace{0.03cm}\ast\,(1)\,a_{1}a_{2}\hspace{0.02cm}b}_{\,{\mathbf k}_{1},\,{\mathbf k}_{2}}
(c^{-\,a_{1}}_{\hspace{0.02cm}{\bf k}_{1}})^{\ast}
(c^{-\,a_{2}}_{\hspace{0.02cm}{\bf k}_{2}})^{\ast}\right]\!
{\mathcal Q}^{-\hspace{0.02cm}c}
+\,\ldots\,.
\]
Two expressions obtained above should be substituted into (\ref{eq:5r}) and (\ref{eq:5t}), respectively, and compared with the asymptotic relations (\ref{eq:4t}) and (\ref{eq:4y}). As a result, we define the first nonzero coefficient function in the representation (\ref{eq:5y})
\begin{equation}
	G^{\,(2)\,a_{1}a_{2}\hspace{0.03cm}b}_{\,{\mathbf k}_{1},\,{\mathbf k}_{2}}
	=
	-\hspace{0.03cm}\frac{i}{2}\,
	\mathscr{T}^{\hspace{0.03cm}(2)\,b\,a_{1}\,a_{2}}_{\; 
		{\mathbf k}_{1},\,{\mathbf k}_{2}}\ \!	2\hspace{0.02cm}\pi\hspace{0.03cm}\delta(\Delta\hspace{0.02cm}\omega_{\hspace{0.03cm}{\mathbf k}_{1},\hspace{0.03cm}{\mathbf k}_{2}})
	\label{eq:5u}
\end{equation}
and therefore, instead of (\ref{eq:5y}) we can now write
\begin{equation}
	\mathcal{T} = 
	\int\!d\hspace{0.02cm}{\bf k}_{1}\hspace{0.02cm} d\hspace{0.02cm}{\bf k}_{2}\, 
	G^{\,(2)\,a_{1}a_{2}\hspace{0.02cm}b}_{\,{\mathbf k}_{1},\,{\mathbf k}_{2}}
	(c^{-\,a_{1}}_{\hspace{0.02cm}{\bf k}_{1}})^{\ast}
	c^{-\,a_{2}}_{\hspace{0.02cm}{\bf k}_{2}}
	{\mathcal Q}^{-\hspace{0.02cm}b}
	+\,\ldots\,.
	\label{eq:5i}
\end{equation}
By virtue of the definition of the function 
$G^{\,(2)\,a_{1}a_{2}\hspace{0.03cm}b}_{\,{\mathbf k}_{1},\,{\mathbf k}_{2}}$ (\ref{eq:5u}) and the property
\[
\mathscr{T}^{\hspace{0.03cm} (2)\hspace{0.03cm}a\,a_{1}\hspace{0.03cm}a_{2}}_{\; {\bf k}_{1},\, {\bf k}_{2}}
=
-\hspace{0.03cm}
\mathscr{T}^{\,\ast\hspace{0.03cm}(2)\,a\,a_{2}\,a_{1}}_{\; {\bf k}_{2},\, {\bf k}_{1}},
\]
for the complete effective amplitude (\ref{eq:2w}), 
which, as we recall, is a consequence of the requirement of reality for the effective Hamiltonian, one can see that the function $\mathcal{T}$ is real, as it should be.\\
\indent In conclusion of this section we note that asymptotic amplitudes 
$c^{\,\pm\,a}_{\hspace{0.02cm}{\bf k}}(t)$, as they were defined in (\ref{eq:3w}), can be expressed through the original amplitudes $c^{\phantom{\hspace{0.03cm}\ast} \!\!a}_{\hspace{0.02cm}{\bf k}}(t)$, $c^{\,\ast\,a}_{\hspace{0.02cm}{\bf k}}(t)$ and the color charge $\mathcal{Q}^{\hspace{0.03cm}a}(t)$. In the leading approximation this relation looks like
\[
c^{\pm\,a}_{\hspace{0.02cm}{\bf k}}(t)
\,=\,
c^{\phantom{\hspace{0.03cm}\ast} \!\!a}_{\hspace{0.02cm}{\bf k}}(t)
+
\frac{i}{2}\int\!d\hspace{0.02cm}{\bf k}_{1}\,
\frac{1}{\Delta\hspace{0.02cm}\omega_{\hspace{0.03cm}
		{\mathbf k},\hspace{0.03cm}{\mathbf k}_{1}}\! \pm i\hspace{0.03cm}0}\,
\mathscr{T}^{\hspace{0.03cm}(2)\,b\,a\,a_{1}}_{\; {\bf k},\, {\bf k}_{1}} c^{\hspace{0.03cm}a_{1}}_{\hspace{0.02cm}{\bf k}_{1}}(t)\hspace{0.03cm}
\mathcal{Q}^{\hspace{0.03cm}b}(t)
\, +\,\ldots\,.
\]

\section{Energy loss of color-charged particle in QCD plasma}
\label{section_6}

As an application of the theory developed in \cite{Markov:2023rlz} and in the previous sections, we study a problem of calculating energy loss of a high-energy color-charged particle traversing a hot quark-gluon plasma, i.e., energy loss due to the scattering process of this particle off soft boson excitations of the medium. As initial expression for energy loss we will use a classical one for energy loss of color-charged particle per unit length being a minimal extension to the color degree of freedom of standard formula for energy loss in an ordinary electromagnetic plasma \cite{Akhiezer:1975}
\begin{equation}
	-\frac{dE}{dx} =
	\frac{1}{\vert{\bf v}\vert}
	\lim\limits_{\tau\rightarrow\infty}
	\frac{1}{\tau}\!\int\limits_{-\tau/2}^{\tau/2}\!\int
	\!d\hspace{0.02cm}{\bf x}\hspace{0.03cm}
	d\hspace{0.02cm}t\int\!d\hspace{0.02cm}\mathcal{Q}_{\hspace{0.02cm}0}
	\,{\rm Re}\,\bigl\langle
	{\bf J}^{\hspace{0.02cm}a}_{\mathcal{Q}}({\bf x},t)\cdot
	{\bf E}^{\hspace{0.02cm}a}_{\hspace{0.03cm}\mathcal{Q}} ({\bf x},t)
	\bigr\rangle
	\label{eq:6q}
\end{equation}
\[
=
\frac{1}{\vert{\bf v}\vert}
\lim\limits_{\tau\rightarrow\infty}
\frac{(2\pi)^4}{\tau}
\int\!d\hspace{0.02cm}{\bf k}\hspace{0.03cm}d\hspace{0.02cm}\omega\!
\int\!d\hspace{0.03cm}\mathcal{Q}_{\hspace{0.02cm}0}\,{\rm Re}\,
\bigl\langle
{\bf J}^{\hspace{0.03cm}\ast\hspace{0.03cm}a}_{\mathcal{Q}}({\bf k},\omega)\cdot{\bf E}^{\hspace{0.02cm}a}_{\hspace{0.03cm}\mathcal{Q}} ({\bf k},\omega)
\bigr\rangle.
\]
Here, chromoelectric field ${\bf E}^{\hspace{0.02cm}a}_{\hspace{0.02cm}\mathcal{Q}}({\bf x},t)$ is one responsible for the particle at the site of its locating. To the procedure of the ensemble average in Eq.\,(\ref{eq:6q}) we have added the integration over the initial value of color charge $\mathcal{Q}^{\hspace{0.03cm}a}_{\hspace{0.02cm}0}$ with a measure that ensures the conservation of the group invariants \cite{PhysRevD.50.4209}
\begin{equation}
	d\hspace{0.02cm}\mathcal{Q}_{\hspace{0.02cm}0} 
	\hspace{0.03cm}\equiv\hspace{0.03cm}
	\mu\hspace{0.03cm}\prod_{e = 1}^{d_A} d\hspace{0.02cm}\mathcal{Q}^{\hspace{0.03cm}e}_{\hspace{0.02cm}0}\,
	\delta(\mathcal{Q}^{\hspace{0.03cm}a}_{\hspace{0.02cm}0}\hspace{0.03cm}
	\mathcal{Q}^{\hspace{0.03cm}a}_{\hspace{0.02cm}0} - q_{2})\,
	\delta(d^{\,a\hspace{0.02cm}b\hspace{0.03cm}c}\hspace{0.03cm}
	\mathcal{Q}^{\hspace{0.03cm}a}_{\hspace{0.02cm}0}\hspace{0.03cm}
	\mathcal{Q}^{\hspace{0.03cm}b}_{\hspace{0.02cm}0}\hspace{0.03cm}
	\mathcal{Q}^{\hspace{0.03cm}c}_{\hspace{0.02cm}0} - q_{3})\,
	\delta(d^{\,a\hspace{0.02cm}b\hspace{0.03cm}c\hspace{0.03cm}d}\hspace{0.03cm}
	\mathcal{Q}^{\hspace{0.03cm}a}_{\hspace{0.02cm}0}\hspace{0.03cm}
	\mathcal{Q}^{\hspace{0.03cm}b}_{\hspace{0.02cm}0}\hspace{0.03cm}
	\mathcal{Q}^{\hspace{0.03cm}c}_{\hspace{0.02cm}0}\hspace{0.03cm} 
	\mathcal{Q}^{\hspace{0.03cm}d}_{\hspace{0.02cm}0} - q_{4})\,
	\dots\,,
	\label{eq:6w}
\end{equation}
where $d_{A} = N^{2}_{c} - 1$ is the dimension of the color Lie algebra $\mathfrak{su}(N_{c})$; $d^{\,a\hspace{0.02cm}b\hspace{0.03cm}c}$ are completely symmetric structure constants of this algebra. All other higher (symmetrized) structure constants for this particular algebra are expressed through
$\delta^{\hspace{0.02cm}a\hspace{0.02cm}b}$ and $d^{\,a\hspace{0.02cm}b\hspace{0.03cm}c}$
(see, for example, \cite{Biedenharn:1963, Klein:1963, Sudbery:1990, Azcarraga:1998}). The number of products of $\delta$-functions on the right-hand side of (\ref{eq:6w}) is equal to the rank of the Lie algebra $\mathfrak{su}(N_{c})$, i.e., $N_{c} - 1$. Thus, in the case of the  $\mathfrak{su}(2_{c})$ algebra we need to keep only the first $\delta$\hspace{0.03cm}-\hspace{0.03cm}function. The constants $q_{2},\,q_{3},\,\ldots$  fix representation-dependent values of the quadratic, cubic, etc., Casimir invariants\footnote{\hspace{0.03cm}In the adjoint representation the group invariant $q_{2}$ is the gluon Casimir $C_{A} = N_{c}$.}. The common multiplier $\mu$ depending on $N_{c}$ in the measure (\ref{eq:6w}) is chosen so that the normalization is valid 
\[
\int d\hspace{0.02cm}\mathcal{Q}_{\hspace{0.02cm}0} = 1,
\] 
the consequence of which, in particular, are the equalities
\begin{equation}
	\int\!d\hspace{0.02cm}\mathcal{Q}_{\hspace{0.02cm}0}\,
	\mathcal{Q}^{\hspace{0.03cm}a}_{\hspace{0.02cm}0}\hspace{0.03cm}
	\mathcal{Q}^{\hspace{0.03cm}b}_{\hspace{0.02cm}0} 
	= 
	\frac{q_{2}}{d_{A}}\,\delta^{\hspace{0.03cm}a\hspace{0.02cm}b},
	\quad
	\int\!d\hspace{0.02cm}\mathcal{Q}_{\hspace{0.02cm}0}\,
	\mathcal{Q}^{\hspace{0.03cm}a}_{\hspace{0.02cm}0}\hspace{0.03cm}
	\mathcal{Q}^{\hspace{0.03cm}b}_{\hspace{0.02cm}0}\hspace{0.03cm}
	\mathcal{Q}^{\hspace{0.03cm}c}_{\hspace{0.02cm}0} 
	= 
	\frac{q_{3}}{d_{A}}
	\biggl(\frac{N^{2}_{c} - 4}{N_{c}}\biggr)^{\!-1}
	d^{\,a\hspace{0.02cm}b\hspace{0.03cm}c}
	\label{eq:6e}
\end{equation}
etc. In addition, we set
\[
\int\!d\hspace{0.02cm}\mathcal{Q}_{\hspace{0.02cm}0}\,
\mathcal{Q}^{\hspace{0.03cm}a}_{\hspace{0.02cm}0} 
= 0.
\]

\section{Effective current}
\label{section_7}

\indent For determining the energy losses we need to know some effective current of a hard color-charged particle in the interaction of the latter with surrounding medium. Here, we again appeal to quantum field theory. In due time, in the framework of $S$-matrix formalism an important notion of {\it radiation operators} was introduced into consideration (see, for example, \cite{Bogolyubov:1975ps, Bogolubov:1990}). Among the radiation operators, the first-order radiation operator plays a special role. This operator (the current operator) is defined by a simple and unified formula:
\begin{equation}
	\hat{J}^{\hspace{0.03cm}(\kappa)l}(x) = -\hspace{0.03cm}i\hspace{0.03cm}\hat{S}^{\hspace{0.03cm}\dagger}\,\frac{\delta\hat{S}}{\delta\hspace{0.03cm}\hat{\phi}^{\, in\hspace{0.03cm}(\kappa)}_{l}(x)}
	\quad
	\mbox{or}
	\quad
	\hat{J}^{\hspace{0.03cm}(\kappa)\hspace{0.03cm}l}(x) = i\,\frac{\delta\hat{S}}{\delta\hspace{0.03cm}\hat{\phi}^{\, out\hspace{0.03cm}(\kappa)}_{l}(x)}\,\hat{S}^{\hspace{0.03cm}\dagger},
	\label{eq:7q}
\end{equation} 
where the index $\kappa$ defines the type of the field $\hat{\phi}^{\hspace{0.03cm}(\kappa)}$. Each of the fields 
$\hat{\phi}^{\hspace{0.03cm}(\kappa)}$ is a tensor-valued or spin-tensor-valued quantity with a finite number of Lorentz 
components $\hat{\phi}^{ \hspace{0.03cm}(\kappa)}_{l},\, (l = 1,\ldots,r_{\kappa})$. This expression, for example for quantum electrodynamics when
$\hat{\phi}_{l}(x)\equiv A_{\mu}(x)$, represents, apart from the sign, the operator of electromagnetic current dressed by radiative corrections.\\
\indent By analogy with quantum field theory, we define the relation between  the classical scattering matrix $\mathcal{S}$ and the effective current of a hard color-charged particle with the help of the following expression:
\begin{equation}
	{\mathcal J}^{\hspace{0.03cm}a\hspace{0.03cm}\mu}_{\mathcal Q}({\bf x},t) 
	= 
	-\hspace{0.03cm}i\hspace{0.03cm}\mathcal{S}^{\hspace{0.03cm}\dagger}
	\hspace{0.03cm}
	\frac{\delta\mathcal{S}}{\delta\hspace{0.03cm}{\mathcal A}^{-\hspace{0.02cm}a}_{\mu}(x)}.
	\label{eq:7w}
\end{equation}
The effective {\it dressed} current (\ref{eq:7w}) of the energetic color particle arises as a result of a screening action of all thermal particles and the interactions with soft color-field excitations of plasma. 
Since the asymptotic in- and out-gauge fields ${\mathcal A}^{-\hspace{0.02cm}a}_{\mu}(x)$ and ${\mathcal A}^{+\hspace{0.02cm}a}_{\mu}(x)$, correspondingly, satisfy free field equations, they can be decomposed into positive and negative frequency parts in an invariant manner valid for all times. Thus we can write, for example,
\begin{equation}
	{\mathcal A}^{-\hspace{0.02cm}a}_{\mu}(x) = \int\!d\hspace{0.02cm}{\bf k}
	\left(\frac{Z_{l}({\bf k})}
	{2\hspace{0.03cm}\omega^{\hspace{0.03cm}l}_{\hspace{0.02cm}{\bf k}}}\right)^{\!\!1/2}\!\!
	\left\{\epsilon^{\hspace{0.03cm}l}_{\mu}({\bf k})\hspace{0.03cm} c^{\phantom{\ast}\!\!-\hspace{0.02cm}a}_{\hspace{0.02cm}{\bf k}}\ \!{\rm e}^{-i\hspace{0.03cm}\omega^{\hspace{0.03cm}l}_{\hspace{0.02cm}{\bf k}}\hspace{0.01cm}t \hspace{0.03cm} +\hspace{0.03cm} i\hspace{0.03cm}{\bf k}\hspace{0.02cm}\cdot\hspace{0.02cm} {\bf x}}
	\,+\,
	\epsilon^{\ast\, l}_{\mu}({\bf k})\, (c^{-\hspace{0.02cm}a}_{\hspace{0.02cm}{\bf k}})^{\ast}\ \!{\rm e}^{\hspace{0.02cm}i\hspace{0.03cm}\omega^{\hspace{0.03cm}l}_{\hspace{0.02cm}{\bf k}}\hspace{0.01cm}t\hspace{0.03cm} -\hspace{0.03cm} i\hspace{0.03cm}{\bf k}\hspace{0.02cm}\cdot\hspace{0.02cm} {\bf x}}
	\right\},
	\label{eq:7e}
\end{equation}
where $c^{\phantom{\ast}\!\!-\hspace{0.02cm}a}_{\hspace{0.02cm}{\bf k}}$ and $(c^{-\hspace{0.02cm}a}_{\hspace{0.02cm}{\bf k}})^{\ast}$ are asymptotic in-amplitudes. An explicit form of the polarization vector of longitudinal mode $\epsilon^{\hspace{0.03cm}l}_{\mu}({\bf k}) = (\epsilon^{\hspace{0.03cm}l}_{0}({\bf k}),	\pmb{\epsilon}^{\hspace{0.03cm}l}({\bf k}))$ in $A_{0}$\hspace{0.03cm}-\hspace{0.03cm}gauge is specified by the following expression:
\begin{equation}
	\epsilon^{l}_{\mu}({\bf k})
	=
	\left.\frac{\tilde{u}_{\mu}(k)}{\sqrt{-\tilde{u}^2(k)}}\ \!\right|_{\rm on-shell}\!,
	\label{eq:7r}
\end{equation}
so we have an evident normalization
\[
\epsilon^{l}_{\mu}({\bf k})\epsilon^{\ast\, l\hspace{0.03cm}\mu}({\bf k}) = -1.
\]
Here, the longitudinal projector $\tilde{u}_{\mu}(k)$ is defined by the first expression in (\ref{ap:A4}). In particular, we have $\tilde{u}_{0}(k) = 0$ in the rest frame of plasma, and as a consequence of the definition (\ref{eq:7r}) we obtain $\epsilon^{\hspace{0.03cm}l}_{0}({\bf k}) = 0$. It is obvious that
\begin{equation}
	(\pmb{\epsilon}^{\hspace{0.03cm}l}({\bf k}))^{\hspace{0.02cm}2}
	= 1
	\quad\mbox{and}\quad
	(\pmb{\epsilon}^{\hspace{0.03cm}l}({\bf k})\cdot\hat{\bf k})
	= 1,
	\label{eq:7t}
\end{equation}
where $\hat{\bf k}\equiv{\bf k}/|{\bf k}|$ and the reality of the polarization vector is taken into account. In the decomposition (\ref{eq:7e}) it is especially important for us the fact that the amplitudes $c^{\phantom{\ast}\!\!-\hspace{0.02cm}a}_{\hspace{0.02cm}{\bf k}}$ and $(c^{-\hspace{0.02cm}a}_{\hspace{0.02cm}{\bf k}})^{\ast}$ are time independent, as will be shown just below.\\
\indent We can invert (\ref{eq:7e}), i.e., express $c^{\phantom{\ast}\!\!-\hspace{0.02cm}a}_{\hspace{0.02cm}{\bf k}}$ and $(c^{-\hspace{0.02cm}a}_{\hspace{0.02cm}{\bf k}})^{\ast}$ in terms of the field function in the coordinate representation ${\mathcal A}^{-\hspace{0.02cm}a}_{\hspace{0.03cm}i}(x),\,i = 1,2,3$, and its time derivative $\dot{\mathcal A}^{-\hspace{0.02cm}a}_{\hspace{0.03cm}i}(x)$ \cite{Bogoliubov:1980, Bjorken:1965zz}. With allowances made for the normalization (\ref{eq:7t}), we derive 
\begin{align}
	&c^{\phantom{\ast}\!\!-\hspace{0.02cm}a}_{\hspace{0.02cm}{\bf k}}
	=
	\frac{1}{2}\left(\frac{2\hspace{0.03cm}\omega^{\hspace{0.03cm}l}_{\hspace{0.02cm}{\bf k}}} {Z_{l}({\bf k})}\right)^{\!\!1/2}\!\!\!
	\int\!\frac{d\hspace{0.02cm}{\bf y}}{(2\pi)^{3}}\,
	e^{\hspace{0.02cm}i\hspace{0.03cm}\omega^{\hspace{0.03cm}l}_{\hspace{0.02cm}{\bf k}}\hspace{0.01cm}t\hspace{0.03cm} -\hspace{0.03cm} i\hspace{0.03cm}{\bf k}\hspace{0.02cm}\cdot\hspace{0.02cm} {\bf y}}\,
	\epsilon^{\hspace{0.03cm}l}_{i}({\bf k})
	\Bigl[{\mathcal A}^{-\hspace{0.02cm}a}_{\hspace{0.03cm}i}({\bf y},t) 
	+ 
	\frac{i}{\omega^{\hspace{0.03cm}l}_{\hspace{0.02cm}{\bf k}}}\,
	\dot{\mathcal A}^{-\hspace{0.02cm}a}_{\hspace{0.03cm}i}({\bf y},t)\Bigr],\notag\\[1ex]
	(&c^{-\hspace{0.02cm}a}_{\hspace{0.02cm}{\bf k}})^{\hspace{0.02cm}\ast}
	=
	\frac{1}{2}\left(\frac{2\hspace{0.03cm}
		\omega^{\hspace{0.03cm}l}_{\hspace{0.02cm}{\bf k}}} {Z_{l}({\bf k})}\right)^{\!\!1/2}\!\!\! 
	\int\!\frac{d\hspace{0.02cm}{\bf y}}{(2\pi)^{3}}\,
	e^{-i\hspace{0.03cm}\omega^{\hspace{0.03cm}l}_{\hspace{0.02cm}{\bf k}}\hspace{0.01cm}t \hspace{0.03cm} +\hspace{0.03cm} i\hspace{0.03cm}{\bf k}\hspace{0.02cm}\cdot\hspace{0.02cm}{\bf y}}\,
	\epsilon^{\hspace{0.03cm}l}_{i}({\bf k})
	\Bigl[{\mathcal A}^{-\hspace{0.02cm}a}_{\hspace{0.03cm}i}({\bf y},t)
	- 
	\frac{i}{\omega^{\hspace{0.03cm}l}_{\hspace{0.02cm}{\bf k}}}\,
	\dot{\mathcal A}^{-\hspace{0.02cm}a}_{\hspace{0.03cm}i}({\bf y},t)\Bigr].
	\notag
\end{align}
As mentioned above, the function $c^{\phantom{\ast}\!\!-\hspace{0.02cm}a}_{\hspace{0.02cm}{\bf k}}$ and $(c^{-\hspace{0.02cm}a}_{\hspace{0.02cm}{\bf k}})^{\hspace{0.02cm}\ast}$ on the left-hand side are time-independent by definition, so the right-hand side of these expressions must also be independent of $t$. For this reason, we can put $t$ equal to an arbitrary constant and, in particular, we can take $t = 0$. Then, instead of the last expressions, we have  
\begin{equation}
	\begin{split}
		&c^{\phantom{\ast}\!\!-\hspace{0.02cm}a}_{\hspace{0.02cm}{\bf k}}
		\,=\,
		\frac{1}{2}\left(\frac{2\hspace{0.03cm}\omega^{\hspace{0.03cm}l}_{\hspace{0.02cm}{\bf k}}} {Z_{l}({\bf k})}\right)^{\!\!1/2}\!\!\!
		\int\!\frac{d\hspace{0.02cm}{\bf y}}{(2\pi)^{3}}\,
		e^{\hspace{0.02cm}-\hspace{0.03cm} i\hspace{0.03cm}{\bf k}\hspace{0.02cm}\cdot\hspace{0.02cm} {\bf y}}\,
		\epsilon^{\hspace{0.03cm}l}_{i}({\bf k})
		\Bigl[{\mathcal A}^{-\hspace{0.02cm}a}_{\hspace{0.03cm}i}({\bf y},0) 
		+ 
		\frac{i}{\omega^{\hspace{0.03cm}l}_{\hspace{0.02cm}{\bf k}}}\,
		\dot{\mathcal A}^{-\hspace{0.02cm}a}_{\hspace{0.03cm}i}({\bf y},0)\Bigr],\\[1ex]
		(&c^{-\hspace{0.02cm}a}_{\hspace{0.02cm}{\bf k}})^{\ast}
		\!=
		\frac{1}{2}\left(\frac{2\hspace{0.03cm}\omega^{\hspace{0.03cm}l}_{\hspace{0.02cm}{\bf k}}} {Z_{l}({\bf k})}\right)^{\!\!1/2}\!\!\! 
		\int\!\frac{d\hspace{0.02cm}{\bf y}}{(2\pi)^{3}}\,
		e^{\hspace{0.03cm} i\hspace{0.03cm}{\bf k}\hspace{0.02cm}\cdot\hspace{0.02cm}{\bf y}}\;
		\epsilon^{\hspace{0.03cm}l}_{i}({\bf k})\,
		\Bigl[{\mathcal A}^{-\hspace{0.02cm}a}_{\hspace{0.03cm}i}({\bf y},0)
		- 
		\frac{i}{\omega^{\hspace{0.03cm}l}_{\hspace{0.02cm}{\bf k}}}\,
		\dot{\mathcal A}^{-\hspace{0.02cm}a}_{\hspace{0.03cm}i}({\bf y},0)\Bigr].
	\end{split}
	\label{eq:7y}
\end{equation}
\indent Next, taking into account the representation (\ref{eq:5e}), we rewrite the right-hand side of the original expression for the effective current (\ref{eq:7w}) in the following form: 
\begin{equation}
	{\mathcal J}^{\hspace{0.03cm}a\hspace{0.03cm}i}_{\mathcal Q}({\bf x},t) 
	= 
	\frac{\delta\hspace{0.03cm}\mathcal{T}}{\delta\hspace{0.03cm}{\mathcal A}^{-a}_{i}(x)}
	=
	\int\! d\hspace{0.02cm}{\bf k}_{1}\!\hspace{0.02cm}
	\left\{\frac{\delta\hspace{0.03cm}{\mathcal T}}{\delta c^{-\phantom{\ast}\!\!a_{1}}_{\hspace{0.02cm}{\bf k}_{1}}}
	\hspace{0.04cm}
	\frac{\delta\hspace{0.01cm}c^{-\phantom{\ast}\!\!a_{1}}_{\hspace{0.02cm}
			{\bf k}_{1}}}{\delta\hspace{0.03cm}{\mathcal A}^{-\hspace{0.02cm}a}_{\hspace{0.02cm}i}(x)}
	\,+\,
	\frac{\!\delta\hspace{0.03cm}{\mathcal T}}
	{\delta (c^{-\ \!\!a_{1}}_{\hspace{0.02cm}{\bf k}_{1}})^{\ast}}
	\hspace{0.03cm}
	\frac{\delta\hspace{0.02cm}(c^{-\ \!\!a_{1}}_{\hspace{0.02cm}{\bf k}_{1}})^{\ast}}
	{\delta\hspace{0.03cm}{\mathcal A}^{-\hspace{0.02cm}a}_{\hspace{0.02cm}i}(x)}\right\}.
	\label{eq:7u}
\end{equation}
With the representation (\ref{eq:7y}) at hand, we easily find the corresponding variational derivatives
\begin{equation}
	\begin{split}
		&\frac{\delta\hspace{0.01cm}c^{-\phantom{\ast}\!\!a_{1}}_{\hspace{0.02cm}
				{\bf k}_{1}}}{\delta\hspace{0.03cm}{\mathcal A}^{-\hspace{0.02cm}a}_{\hspace{0.02cm}i}(x)}
		=
		\delta^{\hspace{0.02cm}a\hspace{0.02cm}a_{1}}
		\frac{1}{2\hspace{0.03cm}(2\pi)^{3}}\,\left(\frac{2\hspace{0.03cm}\omega^{\hspace{0.03cm}l}_{\hspace{0.02cm}{\bf k}_{1}}} {Z_{l}({\bf k}_{1})}\right)^{\!\!1/2}\!\!
		\,
		e^{\hspace{0.02cm} - \hspace{0.03cm} i\hspace{0.03cm}{\bf k}_{1}\hspace{0.02cm}\cdot\hspace{0.02cm} {\bf x}}\,
		\epsilon^{\hspace{0.03cm}l}_{i}({\bf k}_{1})\hspace{0.03cm}\delta(t),\\
		&\frac{\delta\hspace{0.02cm}(c^{-\ \!\!a_{1}}_{\hspace{0.02cm}{\bf k}_{1}})^{\ast}}
		{\delta\hspace{0.03cm}{\mathcal A}^{-\hspace{0.02cm}a}_{\hspace{0.02cm}i}(x)}
		=
		\delta^{\hspace{0.02cm}a\hspace{0.02cm}a_{1}}
		\frac{1}{2\hspace{0.03cm}(2\pi)^{3}}\,\left(\frac{2\hspace{0.03cm}
			\omega^{\hspace{0.03cm}l}_{\hspace{0.02cm}{\bf k}_{1}}} 
		{Z_{l}({\bf k}_{1})}\right)^{\!\!1/2}\!\!
		\,
		e^{\hspace{0.03cm}i\hspace{0.03cm}{\bf k}_{1}\hspace{0.02cm}\cdot\hspace{0.02cm} {\bf x}}\,
		\epsilon^{\hspace{0.03cm}l}_{i}({\bf k}_{1})\hspace{0.03cm}\delta(t).
	\end{split}
	\label{eq:7i}
\end{equation}
In deriving these relations we have assumed the functional derivative of the gauge potential with derivative $\dot{\mathcal A}^{-\hspace{0.02cm}a}_{\hspace{0.03cm}i}({\bf y},0)$ with respect to ${\mathcal A}^{-\hspace{0.02cm}a}_{\hspace{0.02cm}i}(x)$ to be zero,  considering that these functions are independent. By using the explicit form for the phase function $\mathcal{T}$, Eq.\,(\ref{eq:5i}), and the variational derivatives (\ref{eq:7i}), we find from (\ref{eq:7u})  the desired effective current vector in the coordinate representation
\[
\pmb{\mathcal J}^{\hspace{0.03cm}a}_{\!{\mathcal Q}}({\bf x},t) 
=\!
\int\!d\hspace{0.02cm}{\bf k}_{1}\hspace{0.02cm} d\hspace{0.02cm}{\bf k}_{2}
\]
\[
\times 
\left\{
G^{\,(2)\,a_{1}\hspace{0.01cm}a\hspace{0.02cm}b}_{\,{\mathbf k}_{1},\,{\mathbf k}_{2}}\hspace{0.01cm}
F^{\phantom{(2)}}_{{\bf k}_{2}\!}\!\!
\pmb{\epsilon}^{\hspace{0.03cm}l}({\bf k}_{2})\,
e^{\hspace{0.02cm} - \hspace{0.03cm} i\hspace{0.03cm}{\bf k}_{2}\hspace{0.02cm}\cdot\hspace{0.03cm}{\bf x}}\,
(c^{-\,a_{1}}_{\hspace{0.02cm}{\bf k}_{1}})^{\ast}
\!+
G^{\,(2)\,a\hspace{0.03cm}a_{2}\hspace{0.02cm}b}_{\,{\mathbf k}_{1},\,{\mathbf k}_{2}}\hspace{0.01cm}
F^{\phantom{(2)}}_{{\bf k}_{1}\!}\!\!
\pmb{\epsilon}^{\hspace{0.03cm}l}({\bf k}_{1})\,
e^{\hspace{0.02cm}i\hspace{0.03cm}{\bf k}_{1}\hspace{0.02cm}\cdot\hspace{0.03cm}{\bf x}}\,
c^{-\,a_{2}}_{\hspace{0.02cm}{\bf k}_{2}}
\right\}\!\hspace{0.02cm}\delta(t)
{\mathcal Q}^{-\hspace{0.02cm}b}.
\]
Here, for the sake of brevity, we have set
\begin{equation}
	F^{\phantom{(2)}}_{{\bf k}}\!\!\!\! 
	\equiv
	\frac{1}{2\hspace{0.03cm}(2\pi)^{3}}\,\left(\frac{2\hspace{0.03cm}
		\omega^{\hspace{0.03cm}l}_{\hspace{0.02cm}{\bf k}}} {Z_{l}({\bf k})}\right)^{\!\!1/2}.	
	\label{eq:7o}
\end{equation}
The corresponding current in the Fourier representation has the form
\begin{align}
&\pmb{\mathcal J}^{\hspace{0.03cm}a}_{\!\! \mathcal Q}({\bf k},\omega)
=
\int\!d\hspace{0.02cm}t\hspace{0.03cm}d\hspace{0.02cm}{\mathbf x}\,
\pmb{\mathcal J}^{\hspace{0.03cm}a}_{\!\!\mathcal Q}
({\mathbf x},t)\, 
e^{\hspace{0.02cm}i\hspace{0.03cm}\omega\hspace{0.03cm}t\hspace{0.03cm} -\hspace{0.03cm}i\hspace{0.03cm}{\bf k}\hspace{0.02cm}\cdot\hspace{0.02cm} {\mathbf x}}
\label{eq:7p}\\[1ex]
&=
(2\pi)^{3}\!\int\!d\hspace{0.02cm}{\bf k}_{1}\, 
G^{\,(2)\,a_{1}\hspace{0.03cm}a\hspace{0.02cm}b}_{\,{\mathbf k}_{1},\,-{\mathbf k}}\hspace{0.01cm}
F^{\phantom{(2)}}_{-{\bf k}\!}
\pmb{\epsilon}^{\hspace{0.03cm}l}(-{\bf k})\,
(c^{-\,a_{1}}_{\hspace{0.02cm}{\bf k}_{1}})^{\ast}
\hspace{0.02cm}{\mathcal Q}^{-\hspace{0.02cm}b}
\notag\\[1ex]
&+
(2\pi)^{3}\!
\int\!d\hspace{0.02cm}{\bf k}_{2}\,
G^{\,(2)\,a\hspace{0.03cm}a_{2}\hspace{0.02cm}b}_{\,{\mathbf k},\,{\mathbf k}_{2}}\hspace{0.01cm}
F^{\phantom{(2)}}_{{\bf k}\!}\!\!\!
\pmb{\epsilon}^{\hspace{0.03cm}l}({\bf k})\,
c^{-\,a_{2}}_{\hspace{0.02cm}{\bf k}_{2}}
{\mathcal Q}^{-\hspace{0.02cm}b}.
\notag	
\end{align}

\section{Final expression for energy loss of hard particle}
\label{section_8}

Now we return to the expression for energy losses (\ref{eq:6q}). The chromoelectric field in (\ref{eq:6q}), caused by the effective current
(\ref{eq:7p}), is defined by the field equation which in the temporal gauge has the form
\[
E^{\hspace{0.03cm}a\hspace{0.03cm}i}_{\mathcal Q}({\bf k},\omega) 
= 
-\hspace{0.03cm}i\hspace{0.03cm}\omega\, 
^\ast\widetilde{\cal D}^{\hspace{0.03cm}i\hspace{0.03cm}j}(k)\hspace{0.03cm}
{\mathcal J}^{\hspace{0.03cm}a\hspace{0.02cm}j}_{\mathcal Q}({\bf k},\omega),
\]
where $i,\,j = 1,2,3$. The nonzero components of the effective gluon propagator, by virtue of the definitions (\ref{ap:A1})\,--\,(\ref{ap:A4}), are given by the following expression:
\begin{equation}
	\,^{\ast}\widetilde{\cal D}^{\hspace{0.03cm}i\hspace{0.02cm}j}(k) =
	\left(\frac{k^2}{\omega^2}\right)
	\frac{{\rm k}^{\hspace{0.03cm}i}\hspace{0.03cm}{\rm k}^{\hspace{0.03cm}j}}{{\bf k}^2}
	\,^{\ast}\!\Delta^{l}(k) 
	+
	\left(\delta^{\hspace{0.03cm}i\hspace{0.03cm}j} - 
	\frac{{\rm k}^{\hspace{0.03cm}i}\hspace{0.03cm}{\rm k}^{\hspace{0.03cm}j}}{{\bf k}^2}\right)\!\!\,^{\ast}\!\Delta^{t}(k).
	\label{eq:8q}
\end{equation}
Substituting the expression for the chromoelectric field $E^{\hspace{0.03cm}a\hspace{0.03cm}i}_{\mathcal Q}(k)$ into Eq.\,(\ref{eq:6q}) and considering the structure of the propagator (\ref{eq:8q}), we lead to the formula for energy loss
\begin{equation}
	-\hspace{0.02cm}\frac{dE}{d\hspace{0.02cm}x} 
	=
	-\hspace{0.02cm}\frac{1}{\vert{\bf v}\vert}
	\,
	\lim\limits_{\tau\rightarrow\infty}
	\frac{(2\pi)^4}{\tau}
	\!\int\!d\hspace{0.02cm}{\bf k}\hspace{0.03cm}d\hspace{0.02cm}\omega\!
	\int\!d\hspace{0.03cm}{\mathcal Q}^{-}
	\,\frac{\!\omega}{{\bf k}^2}\,\biggl\{
	\frac{k^2}{{\omega}^2}
	\,\bigl\langle\vert({\bf k}\cdot
	\pmb{\mathcal J}^{\hspace{0.03cm}a}_{\!\! \mathcal Q}({\bf k},\omega))\vert^{\hspace{0.03cm}2}\hspace{0.03cm}
	\bigr\rangle\,{\rm Im}(\hspace{0.03cm}^{\ast}{\!\Delta}^{l}(k))
	\label{eq:8w}
\end{equation}
\[
+\,
\bigl\langle\vert({\bf k}\times
\pmb{\mathcal J}^{\hspace{0.03cm}a}_{\mathcal Q}({\bf k},\omega))\vert^{\hspace{0.03cm}2}\hspace{0.03cm}
\bigr\rangle\,{\rm Im}(\hspace{0.03cm}^{\ast}{\!\Delta}^t(k))\biggr\},
\]
where now the integration measure $d\hspace{0.01cm}{\mathcal Q}^{-}$  is defined for the asymptotic value of the color charge ${\mathcal Q}^{-\hspace{0.02cm}a}$. We are interested in the contribution to energy loss caused by scattering off the longitudinal plasma waves, which is proportional to ${\rm Im}\,(\!\,^{\ast}{\!\Delta}^l(p))$. By using the Fourier transform
$\pmb{\mathcal J}^{\hspace{0.03cm}a}_{\!\! \mathcal Q}({\bf k},\omega)$, Eq.\,(\ref{eq:7p}), and the last equality in (\ref{eq:7t}), we reduce the correlation function in the integrand (\ref{eq:8w}) to the following expression:
\begin{equation}
	\bigl\langle\vert({\bf k}\cdot
	\pmb{\mathcal J}^{\hspace{0.03cm}a}_{\!\! \mathcal Q}({\bf k},\omega))\vert^{\hspace{0.03cm}2}\hspace{0.03cm}
	\bigr\rangle
	\vspace{-0.2cm}
\label{eq:8e}
\end{equation}
\begin{align}
	=
	(2\pi)^{6}\biggl\{&F^{\,2}_{-{\bf k}}\hspace{0.03cm}{\bf k}^{2}
	\!\int\!d\hspace{0.02cm}{\bf k}_{1}\hspace{0.03cm}
	d\hspace{0.02cm}{\bf k}^{\prime}_{1}\, 
	G^{\,(2)\,a_{1}\hspace{0.03cm}a\hspace{0.03cm}b}_{\,{\mathbf k}_{1},\,-{\mathbf k}}\,
	G^{\,\ast\,(2)\,a^{\prime}_{1}\hspace{0.03cm}a\hspace{0.03cm}b^{\hspace{0.03cm}\prime}}_{\,{\mathbf k}^{\prime}_{1},\,-{\mathbf k}}
	\bigl\langle(c^{-\,a_{1}}_{\hspace{0.02cm}{\bf k}_{1}})^{\ast}c^{-\,a^{\prime}_{1}}_{\hspace{0.02cm}{\bf k}^{\prime}_{1}}
	\bigr\rangle
	\notag\\[1ex]
	+\,
	&F^{\,2}_{{\bf k}}\hspace{0.03cm}{\bf k}^{2}
	\!\int\!d\hspace{0.02cm}{\bf k}_{2}\hspace{0.03cm}
	d\hspace{0.02cm}{\bf k}^{\prime}_{2}\, 
	G^{\,(2)\,a\hspace{0.03cm}a_{2}\hspace{0.03cm}b}_{\,{\mathbf k},\,{\mathbf k}_{2}}\,
	G^{\,\ast\,(2)\,a\hspace{0.03cm}a^{\prime}_{2}\hspace{0.03cm}b^{\hspace{0.03cm}\prime}}_{\,
		{\mathbf k},\,{\mathbf k}^{\prime}_{2}}
	\bigl\langle(c^{-\,a^{\prime}_{2}}_{\hspace{0.02cm}{\bf k}^{\prime}_{2}})^{\ast}c^{-\,a_{2}}_{\hspace{0.02cm}{\bf k}_{2}}
	\bigr\rangle\biggr\}\hspace{0.03cm}
	{\mathcal Q}^{-\hspace{0.02cm}b}
	{\mathcal Q}^{-\hspace{0.02cm}b^{\hspace{0.03cm}\prime}}.
	\notag
\end{align}
Here on the right-hand side, we have left only terms with non-trivial correlation functions, which we represent as usual
\begin{equation}
\bigl\langle(c^{-\,a^{\phantom{\prime}}_{1}}_{\hspace{0.02cm}{\bf k}_{1}})^{\ast}c^{-\,a^{\prime}_{1}}_{\hspace{0.02cm}{\bf k}^{\prime}_{1}}
\bigr\rangle
=
{\mathcal N}^{\hspace{0.03cm}-\hspace{0.02cm}a^{\phantom{\prime}}_{1} a^{\prime}_{1}}_{{\bf k}_{1}}\hspace{0.03cm}
\delta({\bf k}_{1} - {\bf k}^{\prime}_{1}),
\quad
\bigl\langle(c^{-\,a^{\prime}_{2}}_{\hspace{0.02cm}{\bf k}^{\prime}_{2}})^{\ast}c^{-\,a^{\phantom{\prime}}_{2}}_{\hspace{0.02cm}
	{\bf k}_{2}}\bigr\rangle
=
{\mathcal N}^{\hspace{0.03cm}-\hspace{0.02cm}a^{\prime}_{2} a^{\phantom{\prime}}_{2}}_{{\bf k}^{\prime}_{2}}\hspace{0.03cm}
\delta({\bf k}^{\prime}_{2} - {\bf k}_{2}).
\label{eq:8ee}
\end{equation}
For the plasmon number density matrix ${\mathcal N}^{\hspace{0.02cm}-\hspace{0.02cm}a\hspace{0.03cm}a^{\prime}}_{\bf k}$ we make use of the color decomposition suggested in \cite{Markov:2023rlz} written in terms of the asymptotic in-variables
\begin{equation}
	{\mathcal N}^{\hspace{0.02cm}-\hspace{0.02cm}a\hspace{0.03cm}a^{\prime}}_{\bf k} 
	= 
	\delta^{\,a\hspace{0.03cm}a^{\prime}}\! 
	N^{-\hspace{0.02cm}l}_{\bf k} 
	\hspace{0.03cm}+\hspace{0.03cm}
	\bigl(T^{\,c}\bigr)^{a\hspace{0.03cm}a^{\prime}}\!
	\mathcal{Q}^{-\hspace{0.03cm}c}\,W^{-\hspace{0.02cm}l}_{\bf k},
	\label{eq:8r}
\end{equation}
where the scalar functions $N^{-\hspace{0.02cm}l}_{\bf k}$ and $W^{-\hspace{0.02cm}l}_{\bf k}$ are the colorless and color parts of the plasmon number density, respectively, and the color generators $T^{\,a}$ in the adjoint representation are defined as $\bigl(T^{\,a}\bigr)^{bc} \equiv -if^{abc}$.\\
\indent Let us analyze first the contribution from the colorless part of the asymptotic plasmon number density, i.e., the contribution proportional to the scalar density $N^{-\hspace{0.03cm}l}_{\bf k}$. Integration of the correlation function (\ref{eq:8e}) over the asymptotic charge ${\mathcal Q}^{-\hspace{0.02cm}a}$ by virtue of (\ref{eq:6e}) gives us 
\[
\int\!d\hspace{0.02cm}\mathcal{Q}^{-}\hspace{0.03cm}
\mathcal{Q}^{-\hspace{0.03cm}b}\hspace{0.03cm}
\mathcal{Q}^{-\hspace{0.03cm}b^{\hspace{0.03cm}\prime}}
= 
\frac{C_{A}}{d_{A}}\,\delta^{\hspace{0.03cm}b\hspace{0.03cm}
	b^{\hspace{0.03cm}\prime}}
\]
and, thus, we can now write down
\begin{equation}
	\int\!d\hspace{0.02cm}\mathcal{Q}^{-}
	\bigl\langle\vert({\bf k}\cdot
	\pmb{\mathcal J}^{\hspace{0.03cm}a}_{\!\! \mathcal Q}({\bf k},\omega))\vert^{\hspace{0.03cm}2}\hspace{0.03cm}
	\bigr\rangle
	=
	(2\pi)^{6}\,\frac{C_{A}}{d_{A}}
	\vspace{-0.2cm}
	\label{eq:8t}
\end{equation}
\[
\times\,
\biggl\{F^{\,2}_{-{\bf k}}\hspace{0.03cm}{\bf k}^{2}
\!\int\!d\hspace{0.02cm}{\bf k}_{1}\, 
G^{\,(2)\,a_{1}\hspace{0.03cm}a\hspace{0.03cm}b}_{\,{\mathbf k}_{1},\,-{\mathbf k}}\,
G^{\,\ast\,(2)\,a_{1}\hspace{0.03cm}a\hspace{0.03cm}b}_{\,
	{\mathbf k}_{1},\,-{\mathbf k}}
N^{-\hspace{0.03cm}l}_{{\bf k}_{1}}
+\,
F^{\,2}_{{\bf k}}\hspace{0.03cm}{\bf k}^{2}
\!\int\!d\hspace{0.02cm}{\bf k}_{1}\, 
G^{\,(2)\,a\hspace{0.03cm}a_{1}\hspace{0.03cm}b}_{\,{\mathbf k},\,{\mathbf k}_{1}}\,
G^{\,\ast\,(2)\,a\hspace{0.03cm}a_{1}\hspace{0.03cm}b}_{\,
	{\mathbf k},\,{\mathbf k}_{1}}
N^{-\hspace{0.03cm}l}_{{\bf k}_{1}}
\biggr\}.
\]
The first term in braces actually doubles the second one with the replacement ${\bf k}\rightarrow - {\bf k}$ in the general expression for energy losses (\ref{eq:8w}). Using the explicit form of the coefficient function $G^{\,(2)\,a_{1}a_{2}\hspace{0.03cm}b}_{\,{\mathbf k}_{1},\,{\mathbf k}_{2}}$, Eq.\,(\ref{eq:5u}), we obtain
\begin{equation}
	G^{\,(2)\,a\hspace{0.03cm}a_{1}\hspace{0.03cm}b}_{\,{\mathbf k},\,{\mathbf k}_{1}}\,
	G^{\,\ast\,(2)\,a\hspace{0.03cm}a_{1}\hspace{0.03cm}b}_{\,
		{\mathbf k},\,{\mathbf k}_{1}}
	=
	\frac{1}{4}\,\mathscr{T}^{\hspace{0.03cm}(2)\,b\,a\,a_{1}}_{\; 
		{\mathbf k},\,{\mathbf k}_{1}}\ \!
	\mathscr{T}^{\,\ast\hspace{0.03cm}(2)\,b\,a\,a_{1}}_{\; 
		{\mathbf k},\,{\mathbf k}_{1}}
	(2\hspace{0.02cm}\pi)^{2}\,[\hspace{0.02cm}\delta(\Delta\hspace{0.02cm}
	\omega_{\hspace{0.03cm}{\mathbf k},\hspace{0.03cm}{\mathbf k}_{1}})\hspace{0.02cm}]^{\hspace{0.03cm}2}.
	\label{eq:8y}
\end{equation} 
By virtue of the color and momentum decomposition of the effective amplitude 
\[
\mathscr{T}^{\hspace{0.03cm}(2)\hspace{0.03cm}a\,a_{1}\hspace{0.03cm} a_{2}}_{\; {\bf k}_{1},\, {\bf k}_{2}} = f^{\hspace{0.03cm}a\,a_{1}\hspace{0.02cm}a_{2}}\,
\mathscr{T}^{\hspace{0.03cm}(2)}_{\; {\bf k}_{1},\, {\bf k}_{2}},
\]
further we have
\[
\mathscr{T}^{\hspace{0.03cm}(2)\,b\,a\,a_{1}}_{\; 
	{\mathbf k},\,{\mathbf k}_{1}}\ \!
\mathscr{T}^{\,\ast\hspace{0.03cm}(2)\,b\,a\,a_{1}}_{\; 
	{\mathbf k},\,{\mathbf k}_{1}}
=
f^{\hspace{0.03cm}b\,a\hspace{0.02cm}a_{1}}\,
f^{\hspace{0.03cm}b\,a\hspace{0.02cm}a_{1}}\,
\bigl|\mathscr{T}^{\hspace{0.03cm}(2)}_{\; {\bf k},\, {\bf k}_{1}}\bigr|^{\hspace{0.03cm}2}
=
N_{c}\hspace{0.03cm}d_{A}\hspace{0.03cm}
\bigl|\mathscr{T}^{\hspace{0.03cm}(2)}_{\; {\bf k},\, {\bf k}_{1}}\bigr|^{\hspace{0.03cm}2}.
\]
By the $\delta$-function squared in (\ref{eq:8y}), we mean as usual \cite{Bjorken:1965zz}
\[
\bigl[\hspace{0.02cm}\delta(\Delta\hspace{0.02cm}\omega_{\hspace{0.03cm}{\mathbf k},\hspace{0.03cm}{\mathbf k}_{1}})\hspace{0.02cm}\bigr]^{2}
=
\frac{1}{2\pi}\,\tau\hspace{0.03cm}\delta(\Delta\hspace{0.02cm}\omega_{\hspace{0.03cm}{\mathbf k},\hspace{0.03cm}{\mathbf k}_{1}}).
\] 
Thus, the product (\ref{eq:8y}) takes the final form
\begin{equation}
	G^{\,(2)\,a\hspace{0.03cm}a_{1}\hspace{0.03cm}b}_{\,{\mathbf k},\,{\mathbf k}_{1}}\,
	G^{\,\ast\,(2)\,a\hspace{0.03cm}a_{1}\hspace{0.03cm}b}_{\,
		{\mathbf k},\,{\mathbf k}_{1}}
	=
	\frac{1}{4}\,\tau\hspace{0.03cm} N_{c}\hspace{0.03cm}d_{A}\hspace{0.03cm}
	\bigl|\mathscr{T}^{\hspace{0.03cm}(2)}_{\; {\bf k},\, {\bf k}_{1}}\bigr|^{\hspace{0.03cm}2}
	(2\pi)\hspace{0.03cm}\delta(\Delta\hspace{0.02cm}\omega_{\hspace{0.03cm}{\mathbf k},\hspace{0.03cm}{\mathbf k}_{1}}).
	\label{eq:8u}
\end{equation} 
Substituting (\ref{eq:8u}) into (\ref{eq:8t}), and then into (\ref{eq:8w}), we arrive at the following expression:
\begin{equation}
	-\hspace{0.03cm}\frac{dE}{dx} 
	=
	-\hspace{0.02cm}\frac{1}{\vert{\bf v}\vert}\,\frac{(2\pi)^{10}}{2}\,
	N^{\hspace{0.02cm}2}_{c}
	\label{eq:8i}
\end{equation}	
\[	
\times\!
\int\!d\hspace{0.02cm}{\bf k}\hspace{0.03cm}
d\hspace{0.02cm}{\bf k}_{1}\hspace{0.03cm}d\hspace{0.02cm}\omega\,
\biggl(\frac{k^2}{{\omega}}\biggr)
F^{\,2}_{{\bf k}}\,
\bigl|\mathscr{T}^{\hspace{0.03cm}(2)}_{\; {\bf k},\, {\bf k}_{1}}\bigr|^{\hspace{0.03cm}2}
N^{-\hspace{0.03cm}l}_{{\bf k}_{1}}\,
(2\pi)\hspace{0.03cm}\delta(\Delta\hspace{0.02cm}\omega_{\hspace{0.03cm}{\mathbf k},\hspace{0.03cm}{\mathbf k}_{1}})
\,{\rm Im}(\hspace{0.03cm}^{\ast}{\!\Delta}^{l}(k)).
\]
\indent As the last step in the integrand on the right-hand side of Eq.\,(\ref{eq:8i}) the following representation for the imaginary part of the scalar longitudinal propagator should be substituted:
\begin{align}
	{\rm Im}\,(^{\ast}{\!\Delta}^l(k))
	&\simeq
	- \pi\hspace{0.04cm}{\rm sign}(\omega)\,
	\delta({\rm Re}\,^{\ast}\!\Delta^{\!-1\,l}(k))
	\notag\\
	&=
	-\pi\hspace{0.04cm}{\rm sign}(\omega)\,
	\biggl(\frac{{\rm Z}_l({\bf k})}{2\hspace{0.03cm}\omega_{\bf k}^l}\biggr)
	[\hspace{0.04cm}\delta(\omega - \omega_{\bf k}^l) +
	\delta(\omega + \omega_{\bf k}^l)\hspace{0.03cm}].
	\notag
\end{align}
The contribution of the second $\delta$\hspace{0.03cm}-\hspace{0.03cm}function in square brackets in fact simply doubles the contribution of the first one. Let us substitute the above representation into (\ref{eq:8i}) and integrate over $\omega$. Recalling the definition of the function $F_{{\bf k}}$, Eq.\,(\ref{eq:7o}), we find the desired expression for energy loss associated with the colorless part of the plasmon number density (\ref{eq:8r}):
\begin{equation}
-\hspace{0.03cm}\frac{dE}{dx} 
=
\frac{1}{\vert{\bf v}\vert}\,\frac{(2\pi)^{6}}{8}\,
N^{\hspace{0.02cm}2}_{c}
\!\int\!d\hspace{0.02cm}{\bf k}\hspace{0.03cm}
d\hspace{0.02cm}{\bf k}_{1}\,
\biggl(\frac{k^2}{\omega_{\bf k}^l}\biggr)
\bigl|\mathscr{T}^{\hspace{0.03cm}(2)}_{\; {\bf k},\, {\bf k}_{1}}\bigr|^{\hspace{0.03cm}2}
N^{-\hspace{0.03cm}l}_{{\bf k}_{1}}\,
\delta(\omega^{\hspace{0.02cm}l}_{\hspace{0.03cm}{\bf k}} - \omega^{\hspace{0.02cm}l}_{\hspace{0.03cm}{\mathbf k}_{1}}
- {\mathbf v}\cdot (\hspace{0.03cm}{\mathbf k} - {\mathbf k}_{1})).
\label{eq:8o}
\end{equation}
\indent It remains for us to perform a similar analysis for the contribution of color part of the plasmon number density proportional to the scalar density $W^{-\hspace{0.03cm}l}_{\bf k}$. With this aim, we return to the intermediate expression (\ref{eq:8e}). To be specific, we consider the integrand in the first term in braces, namely:
\[
G^{\,(2)\,a_{1}\hspace{0.03cm}a\hspace{0.03cm}b}_{\,{\mathbf k}_{1},\,-{\mathbf k}}\,
G^{\,\ast\,(2)\,a^{\prime}_{1}\hspace{0.03cm}a\hspace{0.03cm}
b^{\hspace{0.03cm}\prime}}_{\,	{\mathbf k}^{\prime}_{1},\,-{\mathbf k}}
\bigl\langle(c^{-\,a_{1}}_{\hspace{0.02cm}{\bf k}_{1}})^{\ast}c^{-\,a^{\prime}_{1}}_{\hspace{0.02cm}{\bf k}^{\prime}_{1}}
\bigr\rangle\hspace{0.03cm}
{\mathcal Q}^{-\hspace{0.02cm}b}
{\mathcal Q}^{-\hspace{0.02cm}b^{\hspace{0.03cm}\prime}}
\hspace{0.6cm}
\]
or 
\begin{equation}
	G^{\,(2)\,a_{1}\hspace{0.03cm}a\hspace{0.03cm}b}_{\,{\mathbf k}_{1},\,-{\mathbf k}}\,
	G^{\,\ast\,(2)\,a^{\prime}_{1}\hspace{0.03cm}a\hspace{0.03cm}
		b^{\hspace{0.03cm}\prime}}_{\,
		{\mathbf k}^{\prime}_{1},\,-{\mathbf k}}
	\bigl(T^{\,c}\bigr)^{a_{1}\hspace{0.03cm}a^{\prime}_{1}}\!
	\,W^{-\hspace{0.03cm}l}_{{\bf k}_{1}}\hspace{0.03cm}
	{\mathcal Q}^{-\hspace{0.02cm}c}
	{\mathcal Q}^{-\hspace{0.02cm}b}
	{\mathcal Q}^{-\hspace{0.02cm}b^{\hspace{0.03cm}\prime}},
	\label{eq:8p}
\end{equation}
if we leave only the pure non-Abelian part in the correlation function $\bigl\langle(c^{-\,a_{1}}_{\hspace{0.02cm}{\bf k}_{1}})^{\ast}c^{-\,a^{\prime}_{1}}_{\hspace{0.02cm}{\bf k}^{\prime}_{1}}
\bigr\rangle$. Here, we will be interested in the overall color factor of this expression (\ref{eq:8p}). To do this, we first need to explicitly write out the color dependence of the functions $G^{\,(2)}$ using the following rule: 
\[
G^{\,(2)\,a_{1}\hspace{0.02cm}a\hspace{0.03cm}b}_{\,{\mathbf k}_{1},\,-{\mathbf k}}
=
f^{\hspace{0.03cm}a_{1}a\hspace{0.03cm}b}\,
G^{\,(2)}_{\,{\mathbf k}_{1},\,-{\mathbf k}},
\qquad
G^{\,\ast\,(2)\,a^{\prime}_{1}\hspace{0.02cm}a\hspace{0.03cm}
	b^{\hspace{0.03cm}\prime}}_{\,
	{\mathbf k}^{\prime}_{1},\,-{\mathbf k}}
=
f^{\hspace{0.03cm}a^{\prime}_{1}a\hspace{0.03cm}b^{\hspace{0.03cm}\prime}}\,
G^{\,\ast\,(2)}_{\,{\mathbf k}^{\prime}_{1},\,-{\mathbf k}},
\] 
and then to integrate over $\mathcal{Q}^{-}$ the symmetric in color indices product of three asymptotic charges in (\ref{eq:8p}). By virtue of relations (\ref{eq:6e}) this integral must be proportional to the totally symmetric structure constants of the color Lie algebra $\mathfrak{su}(N_{c})$, i.e.,
\[
\int\!d\hspace{0.02cm}\mathcal{Q}^{-}\hspace{0.03cm}
{\mathcal Q}^{-\hspace{0.02cm}c}
{\mathcal Q}^{-\hspace{0.02cm}b}
{\mathcal Q}^{-\hspace{0.02cm}b^{\hspace{0.03cm}\prime}}
\sim\, 
d^{\,c\,b\,b^{\hspace{0.03cm}\prime}}.
\]
It is not difficult to see that, as a result, the color factor in the expression (\ref{eq:8p}) is proportional to the trace of the product of four generators:
\[
{\rm tr}\hspace{0.03cm}\bigl(T^{\,a}\hspace{0.02cm} T^{\,c}\hspace{0.02cm} T^{\,a}\hspace{0.02cm}D^{\,c}\hspace{0.03cm}\bigr)
=
\frac{1}{2}\,N_{c}\,
{\rm tr}\hspace{0.03cm}\bigl(T^{\,c}\hspace{0.02cm}D^{\,c}\hspace{0.03cm}\bigr)
= 0.
\]
Here, we introduce a matrix $D^{a}$ (in addition to $T^{a}$) with components $\bigl(D^{\,a}\bigr)^{\hspace{0.01cm}b\hspace{0.03cm}c}\equiv 
d^{\hspace{0.03cm}a\hspace{0.02cm}b\hspace{0.03cm}c}$ and use the relation 
\[
T^{\,a}\hspace{0.02cm}T^{\,b}\hspace{0.03cm}T^{\,a} = \frac{1}{2}\,N_{c}\hspace{0.03cm}T^{\,b}.
\]
Thus, the contribution to energy loss associated with color part of the plasmon number density is zero. The reason for this lies in the fact that the color factor of this contribution, which is not related to the dynamics of the system, vanishes.

\section{Estimate of energy loss}
\label{section_9}

\begin{figure}[hbtp]
	\vspace{-0.2cm}
	\begin{center}
		\includegraphics[width=1\textwidth]{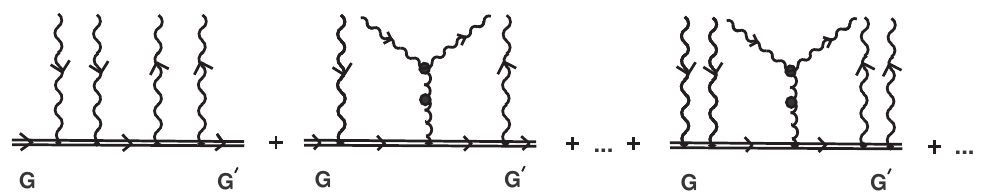}
	\end{center}
	\vspace{-0.2cm}
	\caption{\small The elastic tree level scattering processes involving four and more plasmons}
	\label{fig3}
	\vspace{-0.3cm}
\end{figure}

With the explicit expression for energy loss (Eq.\,(\ref{eq:8o})) at hand, now we can roughly estimate $(-{dE}/{dx})$ at the order-of-magnitude level. First of all we estimate an order of the effective amplitude $\mathscr{T}^{\hspace{0.03cm}(2)}_{\; {\bf k},\, {\bf k}_{1}}$, Eq.\,(\ref{eq:2w}). In the soft region of the momentum scale, when
\[
|{\bf k}| \sim g\hspace{0.02cm}T,\quad 
\omega^{\hspace{0.02cm}l}_{\hspace{0.03cm}{\bf k}} \sim g\hspace{0.02cm}T, 
\]
we have for the eikonal propagator $1/v \cdot k$, the plasmon-hard particle vertex ${\upphi}^{\phantom{\ast}}_{\hspace{0.03cm}{\bf k}}$ and the three-plasmon vertex ${\mathcal V}_{\, {\bf k}, {\bf k}_{1\!}, {\bf k}_{2}}$ based on their definitions (\ref{eq:2ww}), (\ref{ap:A5}), the following estimates
\[
\frac{1}
{\omega^{\hspace{0.03cm}l}_{\hspace{0.03cm}{\bf k}} - {\bf v}\cdot {\bf k}} \sim \frac{1}{g\hspace{0.02cm}T},
\qquad
{\upphi}^{\phantom{\ast}}_{\hspace{0.03cm}{\bf k}}
\sim \frac{1}{(g\hspace{0.02cm}T)^{1/2}},
\qquad
{\mathcal V}_{\, {\bf k}, {\bf k}_{1\!}, {\bf k}_{2}}
\sim \frac{1}{(g\hspace{0.02cm}T)^{1/2}}.
\]
Taking into account these expressions, we obtain from (\ref{eq:2w})
\[
\mathscr{T}^{\hspace{0.03cm}(2)}_{\; {\bf k},\, {\bf k}_{1}}
\sim \frac{1}{T^{\hspace{0.02cm}2}}. 
\]
Further, the integration measure in (\ref{eq:8o}) has an estimate
\[
d\hspace{0.02cm}{\bf k}\hspace{0.03cm}d\hspace{0.02cm}{\bf k}_{1}
\hspace{0.03cm}
\delta(\omega^{\hspace{0.02cm}l}_{\hspace{0.03cm}{\bf k}} - \omega^{\hspace{0.02cm}l}_{\hspace{0.03cm}{\mathbf k}_{1}}
- {\mathbf v}\cdot (\hspace{0.03cm}{\mathbf k} - {\mathbf k}_{1}))
\sim (g\hspace{0.02cm}T)^{5}.
\]
Considering all the above, we find a rough estimate for the energy loss (\ref{eq:8o})
\[
-\hspace{0.03cm}\frac{dE}{dx} 
\sim N^{\hspace{0.02cm}2}_{c} g^{6}\hspace{0.03cm}T^{\hspace{0.02cm}2} N^{-\hspace{0.03cm}l}_{{\bf k}}. 
\]
If we now set for the asymptotic plasmon number density\footnote{\hspace{0.03cm}This estimate is a consequence of the definition of the normal bosonic field variables $c^{\phantom{\ast}\!\!-\hspace{0.02cm}a}_{\hspace{0.02cm}{\bf k}}$ and $(c^{-\hspace{0.02cm}a}_{\hspace{0.02cm}{\bf k}})^{\ast}$, Eq.\,(\ref{eq:7y}), of the definition of the correlation function (\ref{eq:8ee}) and of the estimates for oscillation amplitude of the asymptotic soft field ${\mathcal A}^{-a}_{i}(x)$ for weakly and highly excited states of QGP (see text below).} $N^{-\hspace{0.03cm}l}_{{\bf k}} \sim 1/g^{\hspace{0.03cm}\rho},\; \rho >0$, then from the last expression  follows
\begin{equation}
-\hspace{0.03cm}\frac{dE}{dx} 
\sim N^{\hspace{0.02cm}2}_{c} g^{\hspace{0.02cm}
6 -\rho}\hspace{0.03cm}T^{\hspace{0.02cm}2}.
\label{eq:9q}
\end{equation}
\indent For a low excited state of the quark-gluon plasma, when $\bigl\vert {\mathcal A}^{-a}_{i}(x)\bigr\vert \sim \sqrt{g}\,T$ (the level of thermal fluctuations at the soft scale \cite{Blaizot:2002, Nauta:2000}), we must set $\rho = 1$ and then
\begin{equation}
\biggl(\!-\hspace{0.02cm}\frac{dE}{dx}\biggr)_{{\rm low}} 
\sim N^{\hspace{0.02cm}2}_{c} g^{\hspace{0.02cm}5}\hspace{0.03cm} T^{\hspace{0.02cm}2}.
\label{eq:9w}
\end{equation}
For a more realistic estimate however, it is necessary to perform an explicit analytical (or numerical) calculation of the double integral over the momenta ${\bf k}$ and ${\bf k}_{1}$ on the right-hand side of (\ref{eq:8o}). The procedure in itself is the subject of a specific study and it is not discussed in the present work. Here, the appearance of logarithmic enhancement of $(-{dE}/{dx})_{\rm low}$ is possible.\\
\indent In the other case of a strong field, when $\bigl\vert{\mathcal A}^{-a}_{i}(x)\bigr\vert \sim T$, we must put $\rho = 2$. Then from the estimate (\ref{eq:9q}) follows
\begin{equation}
\biggl(\!-\hspace{0.02cm}\frac{dE}{dx}\biggr)_{{\rm high}} 
\sim N^{\hspace{0.02cm}2}_{c} (g^{\hspace{0.02cm}2}\hspace{0.02cm} T)^{\hspace{0.02cm}2}.
\label{eq:9e}
\end{equation}
When the system under consideration is highly excited, we can expect that higher order scattering processes (see Fig.\ref{fig3}) 
compared to the scattering processes presented in Fig.\,\ref{fig1}, become of the same order in magnitude, and the problem of resummation of all relevant contribution arises.\\
\indent As mentioned in Introduction, the energy losses (47) due to the elastic scattering processes of a hard particle off the soft collective excitations of the quark-gluon plasma are of academic interest in contrast to radiative and collision losses. This can be clearly seen from the energy loss estimate (50). This estimate is suppressed by the strong coupling constant $g$ compared to similar estimates obtained earlier in the framework of perturbative QCD-based transport models or semiclassical frameworks for energy losses due to the medium-induced gluon bremsstrahlung (see, for example, \cite{Gyulassy:1994, Wang:1995, Baier:1997, Baier:2000, Gyulassy:2001, Zakharov:2001, Zakharov:2007, Qin:2008, Djordjevic:2009}) and the elastic collisions with hard thermal particles \cite{Bjorken:1982, Thoma:1991, Braaten:1991, Baier:2000, Adil:2007, Wang:2007, Qin:2008}.\\
\indent However, this suppression occurs when the QGP state is close to thermal equilibrium. In the case when the state of the system is far from equilibrium, due to the estimation (51), the energy loss is comparable in the magnitude with radiative and collision losses. This is consequence of the large values of the soft gluon occupation number.

\section{Conclusion}
\setcounter{equation}{0}
\label{section_10}

In this paper, using the Hamilton equations for the normal bosonic field variable and the color charge of a hard particle, the classical scattering matrix for the process of elastic scattering of the hard color particle off soft bosonic excitations of the quark-gluon plasma has been determined. For this purpose, we have used the Zakharov-Schulman approach developed in the formalization of description of the so-called Hamiltonian wave systems of various physical nature. Sufficient universality of this approach allowed us to propose a method of constructing a classical $\mathcal{S}$-matrix for such a complex object as an essentially nonequilibrium quark-gluon plasma interacting with ultrarelativistic color-charged partons injected from the outside as a result of hard collisions of strongly interacting particles. On the basis of the obtained classical scattering matrix, the effective color current generating this interaction process was found, that in turn allowed us to determine an expression for energy loss of the fast color-charged particle with integer spin, moving in the high-temperature non-Abelian plasma.\\
%
%
\indent A generalization of the results obtained in this study to the fermionic sector of hard and soft excitations in a quark-gluon plasma is of significant theoretical and practical interest.
Note that the consideration of scattering processes with a change of statistics of soft and hard modes appear to be rather complicated already in the very attempt to write out a mathematical apparatus that adequately addresses this problem (see, for example, \cite{Markov:2006wp, Markov:2008wz}). Here, to describe the color degrees of freedom of both the hard color-charged particle with half-integer spin and the soft normal fermionic field variables, it is suggested to use functions taking values in Grassmann algebra. A systematic discussion of the application of elements of this algebra in the framework of physical field theories, as well as theories with higher derivatives, can be found, for example, in the monograph \cite{Gitman:1990}. In constructing a general Hamiltonian wave theory of QGP including bosonic and fermionic (hard and soft) degrees of freedom it will be necessary to construct a generalized nonlinear system of dynamical equations of the Wong type describing the evolution of both the ordinary (commutative) classical color charge and the color charges of Grassmann nature in external stochastic gauge and fermionic fields. Here, it will also be necessary to generalize the construction of the corresponding canonical transformations, which include simultaneously bosonic and fermionic degrees of freedom of the collective excitations of the quark-gluon plasma, and the degrees of freedom associated with the commutative charge $\mathcal{Q}^{\hspace{0.02cm}a}$ and with the Grassmann color charges ${\uptheta}^{\,\ast\ \!\!i}$ and ${\uptheta}^{\phantom{\ast}\!\!i},\,i = 1,\ldots,N_{c}$, of hard test particles with integer and half-integer spins. Additionally, it will be necessary to determine the canonicity conditions for these transformations.\\
\indent However, we can already now say a few words about some of the technical aspects of this extension such as energy losses. The general definition for the first-order radiation operators (\ref{eq:7q}) allows, by analogy with the effective bosonic current (\ref{eq:7w}), to write out the effective fermionic current defined by the classical scattering matrix 
\[
\upeta^{\hspace{0.03cm}i}_{\hspace{0.03cm}\alpha}({\bf x},t) 
= 
-\hspace{0.03cm}i\hspace{0.03cm}\mathcal{S}^{\hspace{0.03cm}\dagger}
\hspace{0.03cm}
\frac{\delta\mathcal{S}}{\delta\hspace{0.03cm}{\bar{\Psi}}^{-i}_{\alpha}(x)},
\]
where ${\Psi}^{-i}_{\alpha}(x)$ is an asymptotic soft fermionic in-field of the system under consideration, obeying the free Dirac equation. In the paper \cite{Markov:2006wp}, the fermionic current $\upeta^{\hspace{0.03cm}i}_{\hspace{0.03cm}\alpha}({\bf x},t)$ was called the fermionic source. Further, as a formula for energy loss in the fermionic sector, we can use the expression proposed in \cite{Markov:2006wp}, namely  
\[
\left(\!-\frac{dE}{dx}\,\right)_{\!{\cal F}}\equiv
\frac{1}{\vert{\bf v}\vert}
\lim\limits_{\tau\rightarrow\infty}
\frac{(2\pi)^4}{\tau}
\sum\limits_{\lambda=\pm}
\int\!d\hspace{0.03cm}{\mathcal Q}^{-}\!
\int\!d\hspace{0.03cm}{\uptheta}^{-}
d\hspace{0.03cm}{\uptheta}^{\ast\hspace{0.01cm}-}
\!\!\int\!q^{0}d\hspace{0.03cm}q^{0}
d\hspace{0.03cm}{\bf q}
\]
\begin{align}
\times\hspace{0.03cm}
\biggl\{&{\rm Im}(^{\ast}{\!\Delta}_{+}(q))\,
\langle\vert\,\bar{u}(\hat{\bf q},\lambda)
\hspace{0.03cm}{\upeta}^{\,i}({\bf v},\chi;{\mathcal Q}^{-},\uptheta^{-}|\,q)
\vert^{\,2\,}\rangle
\notag\\[1ex]
+\,
&{\rm Im}(^{\ast}{\!\Delta}_{-}(q))\,
\langle\vert\,\bar{v}(\hat{\bf q},\lambda)
\hspace{0.03cm}{\upeta}^{\,i}({\bf v},\chi;{\mathcal Q}^{-},\uptheta^{-}|\,q)
\vert^{\,2\,}\rangle\biggr\}.
\notag	
\end{align}
Here, $^{\ast}{\!\Delta}_{\pm}(q)$ represent the scalar quark propagators, the poles of which determine the normal and abnormal (plasmino) modes of oscillations in the fermion sector of the collective excitations of the QGP \cite{Markov:2021nhe}.    
This formula is complementary to the formula (\ref{eq:8w}). The fermionic current $\upeta^{\hspace{0.03cm}i}$ is in general a complicated function depending on the velocity of a hard particle ${\bf v}$, a spinor $\chi$ describing its polarization state and asymptotic color charges: the usual commutative charge ${\mathcal Q}^{-\hspace{0.02cm}a}$, the Grassmann charge ${\uptheta}^{-\hspace{0.02cm}i}$, and its conjugate. The Grassmann color charges belong to the fundamental representation of the $SU(N_{c})$ group.\\
\indent Thus, the whole construction finally reduces to the determination of the corresponding classical scattering matrix for the scattering processes involving hard and soft Bose- and Fermi-excitations of QGP. This $\mathcal{S}$-matrix is determined according to the same scheme outlined in Sections \ref{section_3}\,-\,\ref{section_5} provided that the corresponding fourth-order effective Hamiltonian ${\mathcal H}^{(4)}$ is known. The computation of this Hamiltonian will be considered in our next paper.

\bmhead{Acknowledgements}
The research of Yu.A. Markov and M.A. Markova was funded by the Ministry of Education and Science of the Russian Federation
within the framework of the project ``Analytical and numerical methods of
mathematical physics in problems of tomography, quantum field theory, and
fluid and gas mechanics'' (no. of state registration: 121041300058-1). D.M.
Gitman would like to thank the S~an Paulo Research Foundation (FAPESP)
and the National Council for Research (CNPq) for support. The work of
N.Yu. Markov was supported by Grant for Postgraduate Students and Young
Employees of Irkutsk State University No. 091-24-303.

\section*{Declarations}


\noindent{\bf Conflict of interest.} The authors have no relevant financial or non-financial interests to disclose.

\begin{appendices}
	
\section{Effective gluon propagator and three-plasmon vertex}\label{secA1}
In this appendix, we give an explicit form of the gluon propagator in the hard temperature loop (HTL) approximation \cite{Blaizot:2002,Braaten:1990}. The expression
\begin{equation}
	^{\ast}\widetilde{\cal D}_{\mu \nu}(k) = - P_{\mu \nu}(k) \,^{\ast}\!\Delta^t(k)
	- \widetilde{Q}_{\mu \nu}(k) \,^{\ast}\!\Delta^l(k)
	- \xi_{0}\ \!\frac{k^{2}}{(k\cdot u)^{2}}\ \!D_{\mu \nu}(k)
	\label{ap:A1}
\end{equation}
is the gluon (retarded) propagator in the $A_{0}$\hspace{0.03cm}-\hspace{0.03cm}gauge that is modified by effects of the medium. Here, the ``scalar'' transverse and longitudinal propagators are defined as:
\begin{equation}
	\hspace{-1cm}\,^{\ast}\!\Delta^{t}(k) = \frac{1}{k^{\hspace{0.03cm}2} - \Pi^{\hspace{0.03cm}t}(k)},
	\qquad\quad\;
	\,^{\ast}\!\Delta^{l}(k) = \frac{1}{k^{\hspace{0.03cm}2} - \Pi^{\hspace{0.03cm}l}(k)},
	\label{ap:A2}
\end{equation}
where
\[
\Pi^{\hspace{0.03cm}t}(k) = \frac{1}{2}\,\Pi^{\mu\nu}(k) P_{\mu\nu}(k),
\qquad
\Pi^{\hspace{0.03cm}l}(k) = \Pi^{\mu\nu}(k)\hspace{0.03cm}\widetilde{Q}_{\mu\nu}(k).
\hspace{0.2cm}
\]
The polarization tensor $\Pi_{\mu \nu}(k)$ in the HTL-approximation has the following form
\[
\Pi^{\mu \nu}(k) = 3\hspace{0.035cm}\omega_{\rm pl}^{2}
\left( u^{\mu}\hspace{0.02cm}u^{\nu} - \omega\!\int\!\frac{d\hspace{0.035cm}\Omega_{\mathbf v}}{4 \pi}
\,\frac{v^{\mu}\hspace{0.02cm}v^{\nu}}{v\cdot k + i\hspace{0.02cm}\epsilon} \right),
\]
where $v^{\mu} = (1,{\bf {\bf v}})$, $k^{\mu} = (\omega, {\bf k})$ is a gluon four-momentum, $d\hspace{0.035cm}\Omega_{\mathbf v}$ is a differential solid angle with respect to the unite vector ${\mathbf v}$ and $\omega_{\rm pl}^2 = g^2(2N_c+N_f)T^2/18$ is the plasma frequency squared. The longitudinal and transverse projectors are defined, respectively, by the following expressions:
\begin{equation}
	\begin{split}
		\widetilde{Q}_{\mu \nu}(k) 
		&=
		\frac{\tilde{u}_{\mu}(k)\hspace{0.03cm}\tilde{u}_{\nu}(k)}{\bar{u}^{\hspace{0.02cm}2}(k)}\, ,\\[1ex]
		P_{\mu\nu}(k) 
		&= 
		g_{\mu\nu} - u_{\mu}\hspace{0.02cm}u_{\nu}
		- \widetilde{Q}_{\mu \nu}(k)\,\frac{(k\cdot u)^{2}}{k^{2}}\,.
		\label{ap:A3}
	\end{split}
\end{equation}
Here, in turn, the four-vectors 
\begin{equation}
	\tilde{u}_{\mu} (k) 
	= 
	\frac{k^{\hspace{0.02cm}2}}{(k\cdot u)}
	\ \! \bigl(k_{\mu} - u_{\mu}\hspace{0.02cm}(k\cdot u)\bigr)
	\quad \mbox{and} \quad
	\bar{u}_{\mu} (k) = k^{\hspace{0.03cm}2}\hspace{0.02cm} 
	u_{\mu} - k_{\mu}\hspace{0.02cm}(k\cdot u)
	\label{ap:A4}
\end{equation}
are projectors on the longitudinal direction of the wave vector written in Lorentz-invariant form in the Hamiltonian and Lorentz gauges, respectively;  $u^{\hspace{0.02cm}\mu}$ is the four-velocity of the medium, which in the rest frame of the plasma has the form $u^{\hspace{0.02cm}\mu}=(1,0,0,0)$.\\
\indent Further, we present an explicit form of the effective three-plasmon vertex function  ${\mathcal V}_{\, {\bf k},\, {\bf k}_{1},\, {\bf k}_{2}}$. It was obtained earlier in \cite{Markov:2020efa} when constructing the Hamiltonian formalism for soft Bose excitations in a hot gluon plasma. This vertex reads
\begin{equation}
{\mathcal V}_{\, {\bf k}, {\bf k}_{1\!}, {\bf k}_{2}} =
\label{ap:A5}
\end{equation}
\[
= 
\frac{g}{2^{3/4}}\!
%
\left(\frac{{\rm Z}_l({\bf k})}{2\omega_{\bf k}^l}\right)^{\!\!1/2}\!\!\!
\frac{\tilde{u}_{\mu}(k)}{\sqrt{\bar{u}^2(k)}}
\prod\limits_{i = 1}^{2}
\left(\frac{{\rm Z}_l({\bf k}_{i})}{2\omega_{\bf k}^l}\right)^{\!\!1/2}\!\!\!
\frac{\tilde{u}_{\mu_{i}}(k_{i})}{\sqrt{\bar{u}^2(k_{i})}}
\,^{\ast}\Gamma^{\mu\mu_1\mu_2}(k,\!- k_{1},\!- k_{2})\Bigr|_{\rm \,on-shell},
\hspace{0.4cm} 
\]
where $\,^{\ast}\Gamma^{\mu\mu_1\mu_2}(k,- k_{1},- k_{2})$ is the effective three-gluon vertex in the HTL-approximation.

\end{appendices}


\end{document}